\documentclass[12pt]{article}
\usepackage{amssymb,graphicx}
\usepackage{graphics}
\usepackage{amsmath,amsfonts}
\def\const{\mbox{const}}
\def\e{{\rm e}}
\def\al{\alpha}

\def\d{\partial}

\def\vk{\varkappa}
\newcommand{\be}{\begin{equation}}
\newcommand{\ee}{\end{equation}}
\newcommand{\bea}{\begin{eqnarray}}
\newcommand{\eea}{\end{eqnarray}}
\newcommand{\bg}{\begin{gather}}
\newcommand{\eg}{\end{gather}}
\newcommand{\bseq}{\begin{subequations}}
\newcommand{\eseq}{\end{subequations}}

\newcommand{\bra}[1]{\langle #1 |}
\newcommand{\ket}[1]{| #1 \rangle}

\newcommand\m{\mu}
\newcommand\D{\Delta}
\newcommand\n{\nu}

\renewcommand{\L}{\Lambda}

\title{
\sc{\huge Emergent Lorentz invariance\\
from Strong Dynamics:\\[2mm] 
Holographic examples}
}
\author{
  Grigory Bednik$^{a}$, Oriol Pujol\`as$^b$ and Sergey Sibiryakov$^{c,d}$
\vspace{.2cm}\\
\normalsize\llap{$^a$}\it Department of Physics and Astronomy,
McMaster University\\ 
\normalsize\it Hamilton, Ontario, Canada L8S 4M1\\
\normalsize\llap{$^b$}
 \it Departament de F\'isica and IFAE,
\normalsize\it Universitat Aut\` onoma de Barcelona,\\
\normalsize\it Bellatera 08193, Barcelona, Spain\\
\normalsize\llap{$^c$}\it Institute for Nuclear Research of the
Russian Academy of Sciences, \\ 
      \normalsize \it  60th October Anniversary Prospect, 7a, 117312
      Moscow, Russia\\
\normalsize\llap{$^d$}\it Physics Department, Moscow State University,
\normalsize \it Vorobjevy Gory, 119991 Moscow, Russia
} 

\begin{document}
\maketitle

\begin{abstract}
We explore the phenomenon of emergent Lorentz invariance in strongly
coupled theories. The strong dynamics is handled using 
the gauge/gravity correspondence. We analyze how the renormalization
group flow towards Lorentz invariance is reflected in the two-point
functions of local operators and in the dispersion relations of the bound
states. The deviations of these observables from the relativistic form
at low energies are found to be power-law suppressed by the ratio of
the infrared and ultraviolet scales. We show that in a certain
subclass of models the velocities of the light bound states stay close
to the emergent `speed of light' even at high energies. We comment on
the implications of our results for particle physics and condensed matter.
\end{abstract}

\section{Introduction}
\label{sec:intro}

Lorentz invariance (LI) is one of the best tested symmetries in nature
\cite{Kostelecky:2008ts}. From this fact 
one usually infers that 
Lorentz symmetry is simply a fundamental property underlying the
particle physics and gravity. 
In this paper, we will entertain the opposite idea: is it possible
that LI is not a fundamental but only an accidental symmetry of the
low energy world ? Can LI be an `emergent symmetry' ?  

One motivation to insist on fundamental physics possibly being
Lorentz-violating (LV) may come from the recently discovered
consistent non-relativistic models of gravity
\cite{Horava:2009uw,Blas:2009qj}, which could by themselves be UV
complete. But we will try not to attach too much to any concrete model
and simply assume that at energies $E>M$ for some high energy scale
$M$ physics is Lorentz violating.
The question that one asks then is: {\em is it possible that the
  deviations from LI become small al low energies ?} If this were to
happen naturally ({\em i.e.}, without fine-tunings of parameters)
it would represent an example of an emergent Lorentz invariance.   

In fact, it has been known for a long time that this phenomenon does
happen quite generically. Nielsen et al. considered in
\cite{Chadha:1982qq,Nielsen:1978is} what happens to a LI quantum field theory when it is
perturbed by LV operators.
Let us denote collectively the couplings in front of these operators
$\kappa_{LV}$. One can then compute the renormalization group (RG)
flow of these couplings and finds that they 
vanish towards the IR. The exercise has been repeated in a
number of contexts and with diverse field contents 
\cite{Iengo:2009ix,Giudice:2010zb,Anber:2011xf} giving always
the same result: the
$\kappa_{LV}$ flow to zero and so in the IR the theory is 
attracted towards LI. 
Moreover, the same phenomenon has been observed in the condensed
matter context 
where, in certain materials, effective LI emerges 
from intrinsically
non-relativistic Hamiltonians 
\cite{Vafek:2002jf,Lee:2002qza,Herbut:2009qb}.

These findings agree with the following general argument 
\cite{riccardo,Sundrum:2011ic}, which shows that
under very broad assumptions LI fixed points are IR attractive and thus
quite generically can serve as the end points of the RG
flows. Consider a LI scale invariant theory (SFT) in $d$ space-time
dimensions. It is believed that if this theory is unitary, it enjoys a
larger symmetry, namely the full group of conformal transformations,
i.e. it is a conformal field theory (CFT), see
\cite{Polchinski:1987dy,Luty:2012ww}. Its deformation away from LI can
be equivalently understood as a perturbation of the original
Lagrangian by primary operators\footnote{Or so-called quasi-primaries
  in the case $d=2$. Descendant operators are total derivatives and
  adding them 
  to the Lagrangian does not have any effect.} 
with free Lorentz indices;
schematically, 
\[
{\cal L}_{CFT}\mapsto {\cal L}_{CFT}+\kappa_{LV}{\cal O}_{\m_1\ldots\m_l}\;.
\]
It is convenient to write the perturbation in the Lorentz-covariant
form by contracting the free indices with appropriately chosen spurion
fields, which can be classified according to the
irreducible representations of the Lorentz group
(cf.~\cite{Bolokhov:2007yc}). Let us for simplicity concentrate on the
case of deformations preserving the symmetry under purely spatial
rotations. Then all relevant spurions can be reduced to the single
time-like vector $u^\m$, such that $u^0=1$, $u^i=0$,
$i=1,2,\ldots,d-1$, and the deformation takes the form,
\be
\label{gendef}
{\cal L}_{CFT}\mapsto {\cal L}_{CFT}+\kappa_{LV}{\cal
  O}_{\m_1\ldots\m_l}u^{\m_1}\ldots u^{\m_l}
\;.
\ee
In other words, it is given by the time-components of traceless
symmetric tensors. Now, in unitary CFT's the dimensions of such
operators are bounded from below \cite{Evans,Mack:1975je,Minwalla:1997ka},
\[
\dim {\cal O}_{\m_1\ldots\m_l}\geq d-2+l\;,
\]
where $l$ is the number of Lorentz indices carried by the operator.
Thus, all operators with more than two indices are automatically
{\em irrelevant}. Among operators with two indices in general there
is only one --- the energy-momentum tensor $T_{\m\n}$ --- which is
exactly marginal, $\dim T_{\m\n}=d$, the others being {\em
  irrelevant}\footnote{Indeed, any marginal symmetric tensor 
operator with two indices
  must be conserved \cite{Mack:1975je,Minwalla:1997ka} and thus
  represents an alternative energy-momentum tensor. This is unlikely
  to happen in an interacting theory.}. 
However, the contribution of $T_{\m\n}$ in (\ref{gendef}) can be
absorbed by the rescaling of the time coordinate \cite{Sundrum:2011ic}.
Thus, the only dangerous
perturbations, i.e. those that can drive the theory away from the LI
fixed point, are related to vector operators with the dimensions in
the window $d-1\leq \dim O_\m\leq d$. If no such operators exist in
the theory, the LI point is stable.
Alternatively,  the vector deformations can be forbidden by a
discrete symmetry, such as T-parity or CPT\footnote{This option is
  realized in the LV Standard Model Extension (SME)
  \cite{Colladay:1998fq} where all dimension-3 LV operators turn out to be
  CPT odd. Therefore, imposing CPT-invariance (which is 
  compatible with LV) eliminates all dimension-3 operators in the SME
  at once. Note that if CPT is not imposed these dim-3 
operators would seriously aggravate the fine-tuning needed to satisfy
the experimental constraints on LV.}.

To avoid misunderstanding, let us stress that the above argument does not
imply that {\em any} RG flow must end up at a LI fixed point. 
This is clearly not true for non-relativistic SFT's with dynamical
critical exponent $z\neq 1$. What this argument suggests instead 
is that the emergence of LI
at low energies is not something exceptional and must be rather
common.

This answers the question formulated at the beginning
of the paper basically in the affirmative. However, 
in practice there is one important qualification. In weakly coupled
theories, there are in general LV operators that are marginal at the
tree level. Important representatives of this class are the operators
describing the differences in the maximal propagation velocities of
various particle species \cite{Coleman:1998ti}, which within the
Standard Model are tightly constrained by experiment. 
Upon acquiring positive anomalous dimensions due to
quantum corrections the coefficients of these operators do flow to zero
in the IR, but too slowly. 
The $\kappa_{LV}$ run
logarithmically with the RG scale and for a fixed amount of running
the suppression of the LV couplings in the IR is not sufficient to fulfill
the experimental bounds.  

There are at least two possible ways to overcome this difficulty: 
\begin{enumerate}

\item[(i)] Endow the field theory with enough symmetries 
in order to reduce the number of allowed
operators of lower dimensions, so that LI becomes an accidental
symmetry of the renormalizable (i.e. composed of operators with
canonical dimensions up to 4) part of the Lagrangian. 
This idea is realized in
the mechanism by  
Groot Nibbelink and Pospelov \cite{GrootNibbelink:2004za}, based on a
non-relativistic form of supersymmetry. Combining the latter with the
requirement that all fields are charged under gauge transformations
removes all renormalizable LV operators and
the theory becomes LI in the IR within weak coupling.  
Some implications of this mechanism have been
explored in \cite{Pujolas:2011sk}. 
On the other hand, the embedding of this mechanism in theories with Lifshitz scaling  $(z\neq1)$ presents certain difficulties \cite{Redigolo:2011bv}.

\item[(ii)] Turn to strong coupling.
Typically, the coefficients of the marginally irrelevant LV operators
run 
according to  
$$
{\frac{d\; \kappa_{LV}}{d \log \mu}} \sim \beta\; g^2\;\kappa_{LV};,
$$
where $g$ represents the coupling constant of the theory and 
$\beta>0$ is a numerical constant. 
If $g$ is classically a marginal coupling it will run
logarithmically (while it remains weak) and this immediately leads to
the log-running for $\kappa_{LV}$. To see what can happen at strong
coupling, assume that the dynamics is such that $g$ flows to a
nontrivial fixed point $g_*$. Then
$\kappa_{LV}$ scales like a power-law --- its running gets strongly
accelerated and the LI IR fixed point is 
approached at a much faster rate\footnote{An alternative way to
  accelerate the running is to increase the number of species
  contributing into the renormalization of $\kappa_{LV}$ 
\cite{Anber:2011xf}.} ! 
In the context of particle physics one can then envisage the following
scenario (cf. \cite{Anber:2011xf}): 
the theory starts non-relativistic at very high energy
scale, it runs into strong coupling at lower energies and becomes
relativistic. Finally, it enters into a confining phase at, say, a
few TeV to give rise to the Standard Model (and, possibly, other
hidden sectors)\footnote{This scenario resembles the `walking
  technicolor' idea \cite{Holdom:1981rm} 
invoked for the solution of the gauge
  hierarchy problem.}.   

\end{enumerate}

In this paper we consider option (ii), refining our original question
as follows: {\em What precise implications does the emergence of LI
  have for
physical observables ? What kind of LV effects survive at low energies
and how suppressed they are ?}

To have a handle on the strong dynamics, we will use the gauge/gravity
correspondence. Namely, we are going to consider strongly
coupled  $d$-dimensional SFT's in the limit of large $N$ (number of
degrees of freedom) and at large 't Hooft coupling, which are believed to
be dual to $(d+1)$-dimensional weakly coupled gravity theories defined
on scale-invariant backgrounds. We are going to consider various
deformations of these SFT's which in particular won't leave them
strictly scale-invariant anymore, but we will continue to refer to the
resulting LV field theories as SFT's for short. 
To study when and how LI can emerge in such LV field theories, we
consider two distinct setups: 
\begin{enumerate}
\item[1)] The SFT is defined as a (Lorentz invariant) CFT coupled to
  another `fundamental' Lorentz Violating sector. The two sectors
  are assumed to be `weakly' coupled at some UV scale, {\em i.e.},
  with a controllably small coupling constant. 
This can be realized holographically as a Randall--Sundrum type model
\cite{Randall:1999ee} 
with an Anti-de-Sitter (AdS)
 bulk geometry and LV boundary conditions in the UV. 
Whether LI emerges (or
  `persists') in the IR in this type of model is already
  nontrivial. The answer depends on the type of coupling between the
  CFT and LV sector, and we will see that LI does emerge
  provided the coupling is relevant.  
 
\item[2)] The SFT is by itself non-relativistic in the UV. To fix ideas,
  we will concentrate on the case when at high energies the SFT
  displays anisotropic scaling characterized by the dynamical
  exponent $z>1$. It is conceivable that such anisotropic SFT's admit
  relevant deformations which generate a flow to a LI fixed
  point in the IR\footnote{An example exists already within free-field theory:
    the single-species free Lagrangian 
$${\cal L}=\dot\phi^2 -
    \frac{\phi(-\Delta)^z\phi}{M^{2z-2}}\;,$$
 where $\Delta\equiv \d_i\d_i$ is the spatial Laplacian and
$z>1$, exhibits anisotropic
    scaling. It admits a deformation by a relevant operator
    $c^2\phi\Delta\phi$, which drives the theory in the IR to be LI 
with an
    emergent speed $c$.}. In these cases, the theory exhibits an
  emergent LI. 
This can be realized holographically as a $d+1$ space-time that
approaches a Lifshitz geometry
\cite{Kachru:2008yh,Taylor:2008tg,Braviner:2011kz} in the UV\footnote{More generally, bulk geometries with Lifshitz scaling may also display hyperscaling violation \cite{Charmousis:2010zz,Gouteraux:2011ce}. For simplicity, we shall not consider this case. }. For a
certain range of $z$, these fixed points do admit relevant
deformations that trigger a flow to a LI fixed point. These
flows are dual to `domain-wall' geometries that interpolate between
Lifshitz in the UV and AdS in the IR.  
\end{enumerate}

In both cases we assess the emergence of LI using two types of
observables. First we analyze the two-point functions of scalar
operators and show that they interpolate from a non-relativistic
behavior at short distances to the LI form at large scales. Second we
explore the dispersion relations of bound states in the confining
phase of the theory, which we model by cutting off the geometry with
an IR brane. This will allow us to study the scaling of the LV effects
in the dispersion relations with the IR cutoff by tracing their
dependence on the brane location.   

In this paper we do not attempt
any applications of our construction to realistic model
building. Rather, our aim is to study the basic 
aspects of the mechanism for emergent LI in strongly coupled
theories.

There are in the literature several works that discuss Lorentz Violation in the context of the Gauge/Gravity correspondence.
The closest to our approach is Ref.\footnote{We thank E.~Kiritsis for bringing this Reference to our attention.} \cite{Gubser:2008gr}, which studies physical observables in the case of a holographic RG  flow between two relativistic fixed points with different limiting speeds.
Our aim, instead, is to study systems that are  genuinely non-relativistic in the UV. Our results qualitatively agree with \cite{Gubser:2008gr} where they overlap.
See also \cite{Kiritsis:2012ta,Gouteraux:2012yr} (and references therein) for
other aspects of Lorentz Violation in strongly coupled theories.

The paper is organized as follows. In Sec.~\ref{sec:RS} we consider the
Randall--Sundrum type model, starting with the analysis of the
correlator in Sec.~\ref{ssec:propel} followed by the study of the
bound state spectrum in Sec.~\ref{ssec:RSbound}. In
Sec.~\ref{sec:flow} we turn to the model of the Lifshitz flow. We
introduce the relevant geometries in Sec.~\ref{sec:geometry}, study
the two-point function of a probe operator in
Sec.~\ref{sec:correlators}
and the bound states in Sec.~\ref{sec:states}. Section
\ref{sec:discussion} is devoted to discussion. Three appendices
contain technical details.

\section{Randall -- Sundrum 
model with Lorentz violation on the UV boundary}
\label{sec:RS}

We consider a slice of $AdS_{d+1}$-space with the metric
\[
ds^2=\bigg(\frac{L}{u}\bigg)^2 (-dt^2+dx_i^2+du^2)\;,
\] 
where $L$ is the AdS radius and $i$ runs from $1$ to $d-1$. The two
boundaries are located at $u_{UV}$ and $u_{IR}\equiv
\Lambda^{-1}$. 
Without loss of
generality we can fix $u_{UV}=L$. In this slice we put a probe 
scalar field
$\phi$ 
with the action
\be
\label{AdSaction}
\begin{split}
S=&\frac{1}{2b}\int d^{d+1}x\sqrt{-\mathrm{g}}\big(-\mathrm{g}^{MN}\d_M\phi\d_N\phi
-\m^2\phi^2\big)\\
&+\frac{1}{2}\int_{u=L}d^d x\;\phi\big(-\d_t^2+c^2(-\D
l^2)\D-\bar\m^2_{UV}\big)\phi
-\frac{(L\L)^d}{2}\int_{u=\Lambda^{-1}}d^dx\;\bar\m_{IR}^2\phi^2\;,
\end{split}
\ee
where we have used that the induced metrics on the branes are
flat. Here $\D\equiv\d_i\d_i$ denotes the spatial Laplacian, and $c^2$
is an arbitrary positive function. When $c^2\neq 1$ the kinetic term
 on the UV brane
explicitly violates the Lorentz invariance (LI). The parameter $l$, with dimensions of length, sets the scale of this violation. 
It is natural to assume it to be of order
$L$, though its 
precise value
will be irrelevant for us. The parameter $b$ will be also assumed to
be of order $L$, unless stated otherwise.
For simplicity, we have not included an induced
kinetic term for $\phi$ on the IR brane. Still, we have included the 
bare mass terms on
both branes, which will allow us to obtain massless modes by an
appropriate tuning of the masses.

Let us work out the $d$-dimensional
dual of the theory (\ref{AdSaction}). The bulk part describes a CFT
containing a scalar operator $O_\phi$ 
with 
dimension\footnote{\label{foot:1}If 
  $0\leq\n<1$ the bulk theory admits an alternative interpretation as
  a CFT with the dimension of the dual operator
$\dim O_\phi=d/2-\n$. The value of the field on the UV brane then
corresponds to the operator itself and the dual action (\ref{Sdual})
is replaced by
\[
S=S_{CFT}+\frac{bL^{1-2\n}}{2}\int d^d x\, O_\phi
\big(-\d_t^2+c^2(-\D
l^2)\D-\bar\m^2_{UV}\big)O_\phi\;.
\]
In what follows we stick to the `standard' interpretation based on
(\ref{dimO}) as being more general and covering also the case
$\n>1$. But the analysis of
Secs.~\ref{ssec:propel}, \ref{ssec:RSbound} applies without changes to
the alternative picture as well.}
\cite{Witten:1998qj} 
\be
\label{dimO}
\dim O_\phi=\frac{d}{2}+\nu\;,
\ee
where
\be
\label{nu}
\nu=\sqrt{\frac{d^2}{4}+(\m L)^2}\;.
\ee
The CFT is deformed both in the IR and the UV. By the standard rules, 
the IR  deformation
corresponds simply to the introduction of a confinement scale $\L$.  
In the UV the CFT is cut off at the scale $L^{-1}$, where it
is coupled to an elementary scalar $\bar\phi(x)$ with Lorentz violating (LV)
action. The latter corresponds to the boundary
value of the scalar field $\phi(x,u)$ in the AdS picture.  
Thus the action of the dual theory has the form,
\be
\label{Sdual}
S=S_{CFT}+\frac{1}{2}\int d^d x\;\bar\phi\big(-\d_t^2+c^2(-\D
l^2)\D-\bar\m^2_{UV}\big)\bar\phi
+\frac{L^{\n-1/2}}{\sqrt{b}}\int d^dx\; \bar\phi\, O_\phi,
\ee
The coefficient in front of the last term deserves an explanation. To
obtain it one first absorbs the factor $1/b$ in front of the bulk
action in (\ref{AdSaction}) by rescaling the field 
$\phi=\sqrt{b}\,\tilde\phi$. Then the interaction of $\tilde{\bar\phi}$ with
the dual CFT does not contain any factors of $b$. Going back to the
canonically normalized four-dimensional field $\bar\phi$ one concludes that
the coupling between $\bar\phi$ and CFT is proportional to
$1/\sqrt{b}$. The dependence on $L$ is then reconstructed on
dimensional grounds. Note that the limit $b\to 0$ corresponds to a
strong interaction between the elementary scalar and the CFT, while
$b\to\infty$, on the contrary, gives weak coupling.   

Importantly,
the interaction between the elementary
scalar and the CFT can be relevant or irrelevant,
depending on the parameter $\nu$. Indeed, the
dimension of the scalar is $d/2-1$, implying that if $0\leq\n<1$
the last term in
(\ref{Sdual}) is relevant, for $\n=1$ it is marginal and for $1<\n$
irrelevant. Correspondingly, the physics is expected to be
quite different in these three
cases. 
The first case ($0\leq\nu<1$) is most interesting for our purposes. In this case 
the interaction between the scalar and the CFT 
is relevant and thus becomes strong in the IR. In other words, the
scalar becomes part of the strongly interacting sector. According to
the discussion in the Introduction, one expects the theory to flow  
to a LI invariant point. We will see that this expectation
is indeed confirmed. The marginal case $\n=1$ is more subtle. We will
see that the theory still flows to a LI fixed point in the IR but the
LV 
corrections die away only logarithmically. Finally, 
for $1<\n$ we expect that the interaction between the
scalar and the CFT becomes less and less important when flowing down to
the IR, so that at low energy we obtain two essentially decoupled sectors: the
{\em Lorentz invariant}
CFT and a {\em Lorentz violating} scalar. This case is studied in
detail in Appendix~\ref{app:irr}, where we show that this expectation
is essentially correct, though the actual picture is somewhat more
subtle. 

\subsection{Propagator of the elementary scalar}
\label{ssec:propel}

To warm up let us consider the setup with the IR brane sent to
infinity which corresponds to the removal of the IR cutoff, $\L\to
0$. The quantity we are interested in is the propagator of the
elementary scalar. To obtain it we fix the value of the scalar field
at the UV boundary, $\phi(x,u=L)=\bar\phi(x)$, and integrate out the
bulk degrees of freedom. Upon performing the Fourier decomposition
along the $d$-dimensional coordinates,
\[
\phi(t,{\bf x},u)\propto \phi_{\omega,{\bf k}}(u)
\e^{-i\omega t +i{\bf kx}}\;,
\]
and rotating to the Euclidean signature the bulk solution reads,
\[
\phi_{w,{\bf k}}(u)=\bigg(\frac{u}{L}\bigg)^{d/2}
\frac{K_\n(p_Eu)}{K_\n(p_EL)}\;\bar\phi_{w,{\bf k}}\;.
\]
Here $w=-i\omega$ is the Euclidean frequency, $p_E=\sqrt{w^2+k^2}$ and
$K_\n(z)$ is the Macdonald function. Substitution of this expression
into (\ref{AdSaction}) yields the quadratic boundary action, from
which one reads off the propagator of the elementary scalar,
\be
\label{fsprop}
{\cal G}_{\bar\phi}(w,{\bf k})\equiv
\langle\bar\phi_{-w,-{\bf k}}\bar\phi_{w,{\bf k}}\rangle
\propto
\bigg[w^2+c^2k^2+\m_{UV}^2+\frac{p_E K_{\n-1}(p_EL)}{bK_\n(p_EL)}\bigg]^{-1}\;,
\ee
where
\be
\label{UVmass}
\m_{UV}^2=\bar\m_{UV}^2-\frac{1}{bL}\bigg(\frac{d}{2}-\n\bigg)\;,
\ee
and we have used the relation,
\[
K'_\n(z)=-K_{\n-1}(z)-\frac{\n}{z}K_\n(z)\;.
\]
For high frequency and momentum, $w,k\gg L^{-1}$, the last term inside
the brackets in (\ref{fsprop}) is linear in $p_E$ and can be neglected
compared to the first two terms. Thus we recover the LV propagator
given by the brane part of the action (\ref{AdSaction}). 

On the other hand, the behavior of the propagator at low momenta is
affected by the interaction of $\bar\phi$ with the CFT. To be able to
probe this regime we will assume that $\m_{UV}$ vanishes, so that
$\bar\phi$ describes a gapless mode\footnote{In fact, it would be
  enough to assume $\m_{UV}\ll L^{-1}$. Setting $\m_{UV}=0$ is just
  convenient to simplify the formulas.}. The analysis is different
depending on whether $\n$ is greater, equal or smaller that $1$. For
$\n>1$ the leading term in the ratio of the Macdonald functions at
$p_EL\ll 1$ is quadratic in $p_E$ and the propagator reads
\be
\label{fsprop1}
{\cal G}_{\bar\phi}^{-1}\propto (1+\vk)w^2+(c^2+\vk) k^2\;,
\ee
where
\be
\label{vk}
\vk=\frac{L}{2b(\n-1)}\;.
\ee
We see that, while the coefficients in front of the frequency and
momentum have received a finite renormalization from the CFT, they are
still different from each other implying that violation of LI
persists down to low energies. This is consistent with the fact that
for $\n>1$ the interaction between the elementary scalar and the CFT
is irrelevant. 

It is worth noting, however, that the CFT contribution grows as we
increase the bare coupling between $\bar\phi$ and the CFT (recall that
the latter coupling is proportional to $b^{-1/2}$). As a result the
$\bar\phi$-propagator is driven closer to the LI form. This
illustrates the general statement made in the Introduction that strong
coupling facilitates the emergence of LI. To avoid confusion we stress
that in this case the strong coupling is not achieved by the RG
running, but simply by setting the bare coupling constant in the UV to
be large. From the AdS perspective this corresponds to increasing the
coefficient in front of the bulk Lagrangian, which in the model
(\ref{AdSaction}) is just a fixed parameter.\footnote{One can envisage
a more general setup where this coefficient is rendered dynamical and
varies along the holographic coordinate $u$. This would
correspond then to the RG running of the coupling which, if it increases
towards IR, would trigger the emergence of LI. We will briefly
come back to this possibility in Sec.~\ref{sec:discussion}.}   

We now turn to the case $\n<1$ when the coupling between $\bar\phi$
and the CFT is relevant. Expanding the Macdonald functions at $p_EL\ll
1$ we obtain,
\be
\label{fsprop2}
{\cal G}_{\bar\phi}^{-1}\propto w^2+c^2k^2+2^{1-2\n}
\frac{(p_EL)^{2\n}}{bL}\cdot\frac{\Gamma(1-\n)}{\Gamma(\n)}
\bigg[1-\frac{\Gamma(1-\n)}{\Gamma(1+\n)}\bigg(\frac{p_EL}{2}\bigg)^{2\n}
\bigg]^{-1}
\;.
\ee
The last term dominates at small momenta\footnote{Note that due to
  the factor in the square brackets this term has a smooth limit at 
$\n\to 0$.}. 
Remarkably, it is LI. Thus
we conclude that, consistently with our previous expectations, the
$\bar\phi$-propagator flows to the LI form in the IR. From
(\ref{fsprop2}) we see that the corrections to LI scale as
$(p_EL)^{2(1-\n)}$, i.e. they are power-law suppressed. Note that the
dimension of $\bar\phi$ at the IR fixed point differs from canonical
and is given by $d/2-\n$. With this renormalized value of dimension
the interaction between $\bar\phi$ and the CFT in (\ref{Sdual})
becomes exactly marginal, as it should be.

The case $\n=1$ requires a separate treatment. In this case the
propagator at small momenta reads,
\be
\label{fsprop3}
{\cal G}_{\bar\phi}^{-1}\propto w^2+c^2k^2+
\frac{p_E^2 L}{b}\bigg(-\log{\frac{p_EL}{2}}-\gamma\bigg)\;,
\ee
where $\gamma$ is the Euler number. The last term still dominates and
thus the IR propagator is LI. However, the LV corrections are only
logarithmically suppressed.

\subsection{The bound state spectrum}
\label{ssec:RSbound}

We now come back to the two-brane model. Introduction of the
confinement scale leads to the appearance of the bound states, which
in the holographic picture correspond to the Kaluza--Klein (KK)
modes. Our goal is to analyze the dispersion relation of these
modes. In this subsection we concentrate on the most interesting cases
$\n<1$ and $\n=1$ where LI emerges at low energies automatically due
to the RG flow. The case $\n>1$ is considered in
Appendix~\ref{app:irr} for completeness.  

The profile of a KK mode with given frequency and momentum is
expressed in terms of the Bessel functions, 
\[
\phi_{\omega,{\bf k}}=A_1 u^{d/2}J_\n(p u)+A_2 u^{d/2} Y_{\n}(pu)\;,
\]
where\footnote{Throughout this subsection we work in the Lorentzian
  signature.} 
\be
\label{p}
p=\sqrt{\omega^2-k^2}
\ee
and $A_1$, $A_2$ are arbitrary constants. At the branes one obtains
the boundary conditions
\bseq
\label{AdSbc}
\begin{align}
\label{AdSbc1}
&\frac{1}{b}{\phi}'_{\omega,{\bf k}}
+\big(\omega^2-k^2c^2(k^2l^2)-\bar\mu_{UV}^2\big)\phi_{\omega,{\bf k}}\bigg|_{u=L}=0\;,\\
\label{AdSbc2}
&-\frac{1}{b}{\phi}'_{\omega,{\bf k}}
-L\Lambda\bar\mu_{IR}^2\phi_{\omega,{\bf k}}\bigg|_{u=\Lambda^{-1}}=0\;.
\end{align}
\eseq
This translates into the system of linear equations for the coefficients
$A_1$, $A_2$ that has non-trivial solutions provided its determinant
vanishes,
\be
\label{AdSdet}
{\bf J}_{UV}{\bf Y}_{IR} -{\bf Y}_{UV}{\bf J}_{IR}=0\;.
\ee
Here we have introduced the notations ($Z$ stands for $J$ or $Y$),
\bseq
\begin{align}
&{\bf
  Z}_{UV}=Z_{\nu-1}(pL)+\frac{(\omega^2-c^2k^2-\mu_{UV}^2)b}{p}Z_{\nu}(pL)\;,
\notag\\
&{\bf Z}_{IR}=
Z_{\nu-1}(p/\L)+\frac{bL\L\mu_{IR}^2}{p}Z_{\nu}(p/\L)\;,\notag
\end{align}
\eseq
where $\m_{UV}$ is given by
(\ref{UVmass}) and
\[
\m_{IR}^2=\bar\m_{IR}^2+\frac{1}{bL}\bigg(\frac{d}{2}-\n\bigg)\;.
\]
The equation (\ref{AdSdet})
determines the dispersion relations of the
KK modes. 

In what follows we restrict to the case $\m_{UV}=\m_{IR}=0$ that
guarantees the existence of a gapless mode in the KK
spectrum. 
Besides, we are interested in the modes with four-momenta
small compared to the UV cutoff, $pL\ll 1$, and thus we can Taylor
expand the corresponding Bessel functions in (\ref{AdSdet}). 
We first consider the
relevant coupling case $0\leq \n<1$. After a straightforward
calculation Eq.~(\ref{AdSdet}) can be cast into the form, 
\be
\label{Jexp1}
\bigg(\frac{p}{2\L}\bigg)^{2\n}
\frac{J_{1-\n}(p/\L)}{\tilde J_{\n-1}(p/\L)}
=\frac{b}{2L}(L\L)^{2(1-\n)}(c^2-1)\frac{k^2}{\L^2}\;,
\ee    
where
\[
\tilde J_{\n-1}(p/\L)=\frac{\Gamma(\n)}{\Gamma(1-\n)}
\bigg[J_{\n-1}(p/\L)+\frac{\Gamma(1-\n)}{\Gamma(1+\n)}
\bigg(\frac{pL}{2}\bigg)^{2\n}J_{1-\n}(p/\L)\bigg]\;.
\]
If $(pL)^{2\n}\ll 1$, which holds automatically for $\n$ not too close
0, the second term in the square brackets can be dropped, so that
$\tilde J_{\n-1}$ becomes simply proportional to $J_{\n-1}$. For any
value of the spatial momentum $k$ equation (\ref{Jexp1}) implicitly
determines $p$ and hence the frequencies. Only the r.h.s. of this
equation violates Lorentz invariance. Note that it is suppressed by
the ratio of the IR and UV cutoffs to a positive power,
$(L\L)^{2(1-\n)}$. In other words, the LV effects vanish as 
a power-law when the IR cutoff is lowered.

\begin{figure}[htb]
\begin{center}
\begin{picture}(270,150)(0,10)
\put(5,0){\includegraphics[scale=0.9]{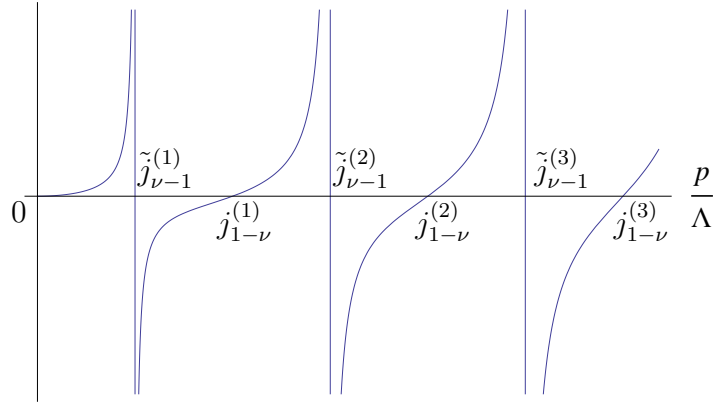}}
\put(0,68){0}
\put(78,64){$j_{1-\n}^{(1)}$}
\put(153,64){$j_{1-\n}^{(2)}$}
\put(228,64){$j_{1-\n}^{(3)}$}
\put(48,84){${\tilde{j}}_{\n-1}^{(1)}$}
\put(122,84){${\tilde j}_{\n-1}^{(2)}$}
\put(198,84){${\tilde j}_{\n-1}^{(3)}$}
\end{picture}
\end{center}
\caption{The l.h.s. of Eq.~(\ref{Jexp1}) as function of $p/\L$
  (qualitative plot).}
\label{Fig:Bessratio}
\end{figure}

To see what happens in more detail, it is useful to plot the l.h.s. of
(\ref{Jexp1}) as a function of $p/\L$. The qualitative shape of this
dependence is shown in Fig.~\ref{Fig:Bessratio}.     
We see that it is composed of an infinite sequence of branches, the
first branch starting from zero and going to $+\infty$, whereas all
other branches vary from $-\infty$ to $+\infty$ as $p/\L$ changes
between the adjacent roots of $\tilde J_{\n-1}$. Each branch corresponds
to a separate KK mode and describes its dispersion relation through
the dependence of the effective mass of the mode $p$ on the spatial
momentum $k$. It is clear from (\ref{Jexp1}) that this leads to the
dispersion relations of the form,
\be
\label{disprel1}
\omega_n^2=k^2+\L^2F_n\big((c^2-1)k^2/k_{LV}^2\big)\;,
\ee
where
\[
k_{LV}=\L (L\L)^{\n-1}\;,
\]
and $F_n$ are dimensionless functions. Note that the scale
$k_{LV}$ lies between the confinement scale $\Lambda$ and the UV
cutoff $L^{-1}$.
 
Let us first analyze the dispersion relations of the massive modes. As
long as
\be
\label{ksmall1}
|c^2-1| k^2\ll k^2_{LV}\;,
\ee   
one can expand the function $F_n$ in (\ref{disprel1}) which yields the
dispersion relation as a power series in momentum,
\be
\label{disprel2}
\omega^2_n=\L^2(j_{1-\n}^{(n)})^2+(1+2\delta c_n)k^2
+\sum_{l\geq 2} \epsilon_l\frac{k^{2l}}{M^{2l-2}_{n,2l}}\;.
\ee
Here $j_{1-\n}^{(n)}$, $n=1,2,\ldots$, is the $n$th positive root of
$J_{1-\n}$ and, assuming that LV in the UV is of order one,
$|c^2-1|\sim 1$, the rest of the parameters are estimated as
\[
\delta c_n\sim (L\L)^{2(1-\n)}~,~~~~~
M_{n,2l}\sim\L(L\L)^{l(\n-1)/(l-1)}\;;
\]
$\epsilon_l=\pm$ stand for the signs of the corresponding terms. 
This expression deserves a few comments. First, we see that the
deviation of the coefficient in front of the $k^2$ term from one is
power-law suppressed by the small factor $(L\L)^{2(1-\n)}$. Second,
contrary to the naive effective field theory logic, the mass
parameters $M_{n,2l}$ suppressing different powers of the momentum are
not equal. This is not surprising as these mass parameters are
generated dynamically from two very different scales: the UV scale of
Lorentz violation $L^{-1}$ and the confinement scale $\L$ governing
the physics of the bound states. While $M_{n,2l}$ are always higher
than $\L$, they can lie below or above $L^{-1}$ depending on the value
of $\n$. For example, the mass scale $M_{n,4}$ appearing in the
quartic term is higher than $L^{-1}$ if $\n<1/2$. 
On the other hand, another statement of the
effective field theory still holds. 
The mere fact that (\ref{disprel2})
is a Taylor expansion implies that, as long as it is valid, the
contribution proportional to $k^{2l}$ is less important than all 
previous terms. In particular, within the validity of (\ref{disprel2})
deviations from LI are dominated by $\delta c_n$.  

When the spatial
momentum reaches $k_{LV}$ the expansion (\ref{disprel2}) breaks down
and the dependence of the frequency on $k$ becomes complicated.  
However, when
$k$ further increases, $F_n$ stabilizes at the nearest root 
${\tilde j}^{(n)}_{\n-1}$ of
$\tilde J_{\n-1}$ (depending on the sign of $c^2-1$ the latter value
is either larger or smaller than $j_{1-\n}^{(n)}$). 
This implies that the dispersion relations
become in this limit relativistic, only with the mass renormalized
compared to the small-momentum regime! Remarkably, the
deviation of the mode velocity from 1 is suppressed at all
momenta\footnote{We omit the region of small momenta where the group
  velocity is different from 1 due to non-zero mass of higher KK
  modes.}, 
\[
\bigg|\frac{d\omega_n}{dk}-1\bigg|\lesssim (L\L)^{2(1-\n)}\;.
\]     
Thus we conclude that for light bound states with masses small
compared to the UV cutoff LV effects are uniformly suppressed at any
value of the spatial momentum or energy. This is interpreted as
follows: the physical scale that determines whether LV is important or
not is neither $k$ nor $\omega$ separately, but rather the effective
particle mass $p$ defined by (\ref{p}).

Now we consider 
the gapless mode. At small momenta $p\ll \L$ the
r.h.s. of (\ref{Jexp1}) is further simplified by expanding the
remaining Bessel functions. This yields a linear
dispersion relation,
\[
\omega^2_0=k^2\bigg[1+\frac{2b(1-\n)}{L}(c^2-1)(L\L)^{2(1-\n)}\bigg]+\ldots\;,
\]
where dots stand for corrections of higher powers in $k^2$ similar to 
(\ref{disprel2}). 
This expression is valid under the condition (\ref{ksmall1}).
 For $k>k_{LV}$ the situation depends
qualitatively on whether $c^2$ is bigger or smaller than 1. In the
former case the gapless mode behaves similar to the massive
ones. Namely, its effective $k$-dependent mass is real and stabilizes to
$p=\L{\tilde j}^{(1)}_{\n-1}$ at high
momenta, see Fig.~\ref{Fig:Bessratio}. Thus LI is approximately
satisfied even by very energetic particles. On the other hand, for
$c^2<1$ the effective mass found from (\ref{Jexp1}) is purely
imaginary. At $k\gg k_{LV}$ the dispersion relation reads,
\[
\omega^2_0=k^2-\frac{4}{L^2}\bigg[\frac{\Gamma(1+\n)}{\Gamma(1-\n)}\cdot
\frac{(1-c^2)bL k^2}{2\n+(1-c^2)bL k^2}\bigg]^{1/\n}\;.
\]
We see that when $k$ approaches the UV cutoff $L^{-1}$ deviations from
LI become of order one. It may even seem that the dispersion relation
turns over and $\omega^2_0$ becomes negative signaling an
instability. However, this is not true: at $k\gtrsim L^{-1}$ 
Eq.~(\ref{Jexp1}) is no longer valid because the condition
$|pL|\ll 1$ is violated. One has to go back to the exact
Eq.~(\ref{AdSdet}). 
A careful analysis shows that at $k\gg L^{-1}$ the
dispersion relation approaches the `bare' non-relativistic 
form $\omega^2_0=c^2 k^2$. We conclude that, within the considered
setup, the `superluminal' case $c^2>1$ leads to more efficient
emergence of LI than the `subluminal' one $c^2<1$. 

For the case $\n=1$, corresponding to the marginal coupling between
the LV scalar and the CFT, the analysis proceeds in essentially the
same way as above. The only difference is the necessity to keep track
of the logarithmic terms in the expansion of the Bessel
functions. This leads to the following
equation for the KK spectrum at $pL\ll 1$,
\[
p^2\bigg[\frac{\pi Y_0(p/\L)}{2J_0(p/\L)}-\log\frac{p}{2\L}
-\gamma+\frac{b}{L}\bigg]
-p^2\log{L\L}
=\frac{b}{L}(c^2-1)k^2\;.
\]
The dependence of the first term
on the l.h.s. on $p$ is similar to that plotted in
Fig.~\ref{Fig:Bessratio}, being again composed of a series of branches
of tg-like shape. The second term involving the large logarithm
$-\log{L\L}$ shifts every branch upward. This implies the following
picture for the massive modes. At small $k$ their dispersion relations
start from
\[
m_n=\L\Big(j_0^{(n)}+O\big(|\log{L\L}|^{-1}\big)\Big)\;.
\]
If the LV scalar is `subluminal', $c^2<1$, the $k$-dependent mass
$p(k)$ always stays close to this value, approaching $\L j_0^{(n)}$ at
$k\to\infty$. On the other hand, in the `superluminal' case $c^2>1$,
the effective mass goes to the next root of the Bessel function,
$p(k)\to\L j_0^{(n+1)}$ at $k\to\infty$. The transition occurrs at
$(c^2-1)k^2\sim k_{LV}^2$ with
\[
k_{LV}=\L\sqrt{-\log{L\L}}\;.
\]   
This is still parametrically higher than $\L$, but the hierarchy is only
logarithmic. The deviations of the mode velocities from 1 are
suppressed by the same logarithm.  

For the gapless mode one again starts from the regime $p\ll\L$ where
one
obtains linear dispersion relation,  
\be
\label{lightmode3}
\omega^2_0=k^2\bigg[1+\frac{c^2-1}{1-(L/b)\log{(L\L)}}\bigg]\;.
\ee
At $\L\to 0$ the velocity of the mode tends to 1 implying emergence of
LI. However, the approach to the LI regime is only logarithmic and
thus too slow to be relevant for realistic particle physics
models. The asymptotics of the dispersion relation at high momenta
$k\gg k_{LV}$ depend on the sign of $c^2-1$, the analysis being
completely
 analogous to the case of relevant coupling. If $c^2>1$ the mode
becomes effectively massive with constant mass $\L j_0^{(1)}$, while
for $c^2<1$ the dispersion relation approaches $\omega^2_0=c^2k^2$.

Let us attempt an intuitive understanding of the surprising result
of this section that the
bound states display the emergent relativistic dispersion relation
even at large momenta $k\gg k_{LV}$.
Actually, we have found this behavior for all KK modes except the zero
mode in the subluminal case $c<1$.
In the 5D picture, the reason behind this observation is
that the purely-LV term on the UV brane $(c^2-1) k^2 \phi^2
\delta(u-u_{UV})$ acts as a $k-$dependent mass localized on the brane. As a
result, at large $k$ and $c>1$ the KK wavefunctions are 
repelled from the brane and the LV effects vanish. For
$c<1$ the effective squared mass is negative and  the zero-mode is
attracted towards the UV brane leading to the strengthening of LV. On
the other hand, the excited KK modes, being orthogonal to the
zero-mode, are still repelled from the brane and their 
$k-$dependent effective mass saturates in the UV at 
the value corresponding to the Dirichlet boundary conditions. 

This has the following CFT interpretation. 
The emergence of LI from
the RG flow means that the theory
possesses a preferred frame which can only be `noticed' by
short-distance probes. 
The excited KK modes map to
resonances --- the states which are composite at all momenta and have
a finite size of the order of the confinement scale 
$\Lambda$. Thus, they simply cannot probe the preferred frame. 
The zero-mode, instead, can be viewed as a mixture of an elementary
and a composite state. 
The size of such an object can be momentum-dependent. 
For $c<1$ the size shrinks with $k$ and the state becomes mostly
elementary. It is now able to probe the UV physics and its velocity
approaches the UV value $c$.
For $c>1$, on the contrary, the admixture of the elementary component
never becomes large and the state remains mostly composite. 
Then just like for the resonances the
emergent relativity persists even for large $k$.

\section{Lifshitz flows}
\label{sec:flow}

While the model considered in the previous section provides a
particularly simple realization of the dynamical emergence of LI
through strong coupling, one may object that the LV sector on the UV
brane is rather ad hoc. Admittedly, it is not straightforward to
couple the
action of the UV brane in 
(\ref{AdSaction}) to the metric, as required for the fully dynamical
formulation\footnote{It is possible that such formulation can be
  achieved by introducing additional degrees of freedom that would
  define a preferred frame on the UV brane, similar to the approach of
\cite{Jacobson:2000xp,Blas:2009yd}.}.  
In this section we study another model that does not suffer from this
drawback, being generally covariant by construction. It was introduced
in \cite{Kachru:2008yh,Taylor:2008tg} and allows to describe RG flows between
non-relativistic fixed points with anisotropic (Lifshitz)
scaling in the UV and LI field theories in the infrared. After 
reviewing the model and the relevant geometries in
Sec.~\ref{sec:geometry}, we address the physical aspects of emergent
LI in Secs.~\ref{sec:correlators}, \ref{sec:states}.

\subsection{The flow geometry}
\label{sec:geometry}

This subsection closely follows the presentation of 
\cite{Braviner:2011kz}. Consider a $(d+1)$-dimensional theory
containing gravity with negative cosmological constant 
and a free massive vector field\footnote{The original model of
  \cite{Kachru:2008yh}, formulated in $d=3$, includes massless 1- and
  2-form fields with a Chern--Simons-type mixing. These are equivalent
to a massive vector on-shell.}. The action reads,
\be
\label{Sgeom}
S=\frac{1}{16\pi G}\int d^{d+1}x\sqrt{-\mathrm{g}}\bigg[R-2\Lambda_c-
\frac{1}{4}{\cal F}_{LN}{\cal F}^{LN}-\frac{M^2}{2}{\cal A}_N {\cal
  A}^N\bigg]\;, 
\ee 
where ${\cal F}_{LN}=\d_L{\cal A}_N-\d_N{\cal A}_L$, as usual.
We look for solutions of this theory that are static and 
invariant under translations and rotations in $(d-1)$ spatial
directions. This leads to the Ansatz,\footnote{The imposed symmetries
  also allow a non-zero value of the ${\cal A}_u$ component. However,
  this vanishes in the
  metric (\ref{metric}) due to the equations of motion for the vector field.}
\be
\label{metric}
ds^2=\bigg(\frac{L}{u}\bigg)^2\Big(-f^2dt^2+\sum_{i=1}^{d-1}dx_i^2
+g^2du^2\Big)\;,~~~~~~
{\cal A}_t=\frac{2}{Mu}fj\;,
\ee
where $f$, $g$, $j$ are functions of the coordinate $u$ to be
determined. It is convenient to 
introduce one more function $h(u)$
parameterizing the field strength,
\[
{\cal F}_{ut}=-\frac{2L fg}{u^2}h\;.
\]
Then the Einstein-Maxwell equations reduce to the set of
first-order ordinary differential equations,
\bseq
\label{veceqs}
\begin{align}
\label{veceq1}
&uh'=-MLgj+(d-1)h\;\\
\label{veceq2}
&uj'=-MLgh-\frac{d-2}{2}j-\frac{jg^2}{d-1}(h^2-j^2+\Lambda_c L^2)\;,\\
\label{veceq3}
&\frac{ug'}{g}=-\frac{g^2}{d-1}(h^2+j^2+\Lambda_c L^2)-\frac{d}{2}\;,\\
\label{veceq4}
&\frac{uf'}{f}=\frac{g^2}{d-1}(h^2-j^2+\Lambda_c L^2)+\frac{d}{2}\;.
\end{align}
\eseq
Note that the first three equations form a closed system for the
functions $g$, $h$, $j$.

Let us investigate the fixed points of this system. These are
characterized by constant values of $g$, $h$, $j$, and
a power-law behavior of the function $f$,
\be
\label{fz}
f=f_0 u^{1-z}\;.
\ee
To understand the physical meaning of these solutions, 
note that 
the corresponding geometries possess an isometry,
\bseq
\label{Lifs}
\begin{align}
\label{Lif1}
&t\mapsto \lambda^z t~,~~~{\bf x}\mapsto \lambda {\bf x}\;,\\
\label{Lif2}
&u\mapsto \lambda u\;,
\end{align}
\eseq 
where $\lambda$ is an arbitrary number. If $z=1$ one recovers the AdS
space-time that by the standard AdS/CFT dictionary is dual to a
relativistic conformal field theory. By the extrapolation of this
logic, the geometries invariant under (\ref{Lifs}) with $z$ different
from 1 have been conjectured \cite{Kachru:2008yh} to be dual to
strongly coupled $d$-dimensional non-relativistic theories invariant
under the anisotropic scaling transformations (\ref{Lif1}). This kind
of scale invariance was considered for the first time in the seminal
works by Lifshitz \cite{Lifshitz}, hence we will refer to the
geometries with the isometries (\ref{Lifs}) as `Lifshitz space-times'.

Setting the l.h.s. of Eqs.~(\ref{veceqs}) to
zero one observes that AdS is always a solution, provided the vector
field vanishes, $h=j=0$. By an appropriate rescaling of coordinates
$f$ and $g$ can be set to 1, which fixes the relation between the AdS
radius and the cosmological constant,
\[
\Lambda_c L^2=-\frac{d(d-1)}{2}\;.
\]
Clearly, this fixed point is Lorent invariant.
Besides, for the vector mass in the range,
\be
\label{ML}
d-1\leq (ML)^2\leq \frac{d(d-1)^2}{3d-4}\;,
\ee
the system (\ref{veceqs}) has two fixed points with Lifshitz scaling,
\bseq
\label{fixedps}
\begin{align}
\label{fixedpsg}
&g_{\pm}^2=\frac{d-1}{2(ML)^2}\bigg[\frac{d(d-1)^2}{(ML)^2}
-d+2\pm\sqrt{\bigg(\frac{d(d-1)^2}{(ML)^2}-d+2\bigg)^2-4(d-1)^2}\bigg]\;,\\
\label{fixedpsj}
&h_{\pm}^2=\frac{d(d-1)}{2}-\frac{(ML)^2}{2}-\frac{(d-1)^2}{2g_{\pm}^2}\;,~~~~~
j_{\pm}^2=\frac{(ML)^2}{2}-\frac{d-1}{2g_{\pm}^2}\;,~~~~
z_{\pm}=\frac{(ML)^2g_{\pm}^2}{d-1}\;,
\end{align}
\eseq
For $(ML)^2< d-1$ only the ``$+$'' branch survives ($j_-^2$ determined
from (\ref{fixedpsj}) becomes negative) and for $(ML)^2>
d(d-1)^2/(3d-4)$ there are no Lifshitz solutions at all. Note that
on the ``$-$'' branch the critical exponent is confined to the
interval  
\be
\label{zinterval}
1\leq z\leq d-1\;,
\ee
while on the ``$+$'' branch it always exceeds\footnote{For a given
  value of $ML$ in the range (\ref{ML}) the critical exponents on the
  two branches are related by a simple formula $z_+z_-=(d-1)^2$.}  
$d-1$. As we are going
to explain shortly, only the ``$-$'' branch is relevant for our
purposes. Thus we will restrict to the choice (\ref{ML}) in what follows.

Our aim is to study RG flows connecting non-relativistic SFTs in the
UV to relativistic CFTs in in the IR. These correspond to the
solutions of the system (\ref{veceqs}) that interpolate between
Lifshitz space-time at small values of the coordinate $u$ and AdS at
$u\to\infty$. To work out if such `domain wall' solutions can be
realized in the model at hand,  
let us analyze the stability properties of the fixed points. 
To
this end we add small perturbations $\delta f(u)$, etc., on top of the
fixed point solutions and expand Eqs.~(\ref{veceqs}) to linear order
in these perturbations.

We start with the AdS fixed point. At  linear order the
perturbations of the metric and the vector field decouple. Consider
first the metric perturbations keeping the vector at zero, $h=j=0$.  
One finds two modes, $\delta f=const\;,~\delta g=0$ and 
$\delta f,~\delta g\propto u^d$. The physical interpretation of these
modes is known: the first corresponds to the source and 
the second to the
expectation value of the energy--momentum tensor of the dual
CFT. 
Indeed,
the growing mode
represents the first term in the expansion of the
AdS-Schwarzschild metric. Thus, the geometry that is obtained by the
excitation of this mode is that of an AdS black hole. In the dual
picture this corresponds to setting the CFT at finite temperature. As in the
present paper we are
interested in the vacuum solutions, we shall set this mode
to zero. On the other hand, the constant mode can be eliminated by the
rescaling of 
the time-coordinate.

Another set of modes involve perturbations of the vector field. These behave
as the power-law,
\[
h\;,~j\propto u^{-\alpha_\pm}\;,~~~~
\alpha_\pm=-\frac{d}{2}\pm\sqrt{\bigg(\frac{d}{2}-1\bigg)^2+(ML)^2}\;.
\]
To have a domain wall solution that asymptotically approaches AdS at
large $u$ we need at least one decaying mode. In other words, the
vector field must be 
heavy enough, so that 
\be
\label{alphamin}
\alpha\equiv\alpha_+>0\;.
\ee 
Note that this condition coincides with the lower bound in
(\ref{ML}). It matches nicely with the
holographic interpretation of the
considered system.
Indeed, the presence of the vector field ${\cal A}_M$ in the bulk
theory implies that the dual LI CFT contains a vector operator 
$O^\m_{\cal A}$, whose dimension is given by the exponent of the growing mode,
\be
\label{dimOA}
\dim O_{\cal A}^\m=d+\al\;.
\ee
The domain wall solution we are looking for corresponds to perturbing
the CFT with the time-component of 
this operator (and in this way breaking LI). For the LI
fixed point to be IR attractive, the operator $O^\m_{\cal A}$ must be
irrelevant, $\dim O_{\cal A}^\m>d$, which is equivalent to
(\ref{alphamin}). 
Note, however, that the upper limit on the allowed masses of the
vector field in (\ref{ML}) implies that $\al$ is bounded from above
 and thus the dimension of $O_{\cal A}^\m$ cannot
deviate parametrically from $d$.
For example, for $d=3,4$ we have that $\al$ is smaller than
$0.13$ and $0.35$ respectively. In fact, $\alpha$ is less than 1 for
all $d\leq 6$.

Let us turn to the perturbations around the Lifshitz fixed
point. The analysis us identical for the ``$+$'' and ``$-$''
branches. Linearizing the first three equations of (\ref{veceqs}) one finds
three modes,
\[
\delta h,~\delta j,~\delta g\propto u^{\beta}\;,
\]
where $\beta$ can take the values,
\begin{align}
\label{beta1}
&\beta_1=z+d-1\;,\\
\label{betapm}
&\beta_{\pm}=\frac{1}{2}\big[z+d-1\pm
\sqrt{(z+d-1)^2+8(z-1)(z-d+1)}\big]\;.
\end{align}
Inserting these into the linearized Eq.~(\ref{veceq4}) we obtain the
corresponding perturbations $\delta f$
that we define as,
\[
f=f_0 u^{1-z}(1+\delta f)\;.
\] 
At this stage we find a  fourth mode $\delta g=\delta h=\delta j=0$,
$\delta f=\const$. In \cite{Braviner:2011kz} this mode together with
the mode corresponding to $\beta_1$ (\ref{beta1}) 
were
interpreted respectively as the source and expectation value of the
energy density 
of the system. Indeed, the energy density operator must be marginal,
i.e. its scaling dimension must be $(z+d-1)$ --- the inverse of the scaling
dimension of the integration measure $dt\, d^{d-1} x$ in the
action\footnote{In our conventions the scaling dimension of the spatial
coordinates is $-1$.} --- which matches with $\beta_1$. 

Consider now the pair of modes (\ref{betapm}). Following the
usual holographic vocabulary one interprets the mode corresponding to
$\beta_-$ as the source of a certain operator $\tilde O_{\cal A}$ 
in the Lifshitz theory,
and the $\beta_+$-mode as its expectation value
\cite{Braviner:2011kz}. Consequently, $\beta_+$ sets the scaling dimension of
the operator. For $1<z<d-1$ we
have $\beta_+< z+d-1$ implying that the operator is relevant,
for $z=d-1$ it is marginal, while for $z>d-1$ it is 
irrelevant. Thus the Lifshitz fixed point can be
UV attractive only in the range (\ref{zinterval}). This excludes from
consideration the ``$+$'' branch of fixed points. The geometrical
counterpart of this statement is that if and only if $1<z<d-1$ both
exponents (\ref{betapm}) are positive implying that all perturbations
decay towards the asymptotic UV boundary $u\to 0$ and the Lifshitz
space-time is UV attractive\footnote{\label{foot:marg}
The situation is somewhat trickier for the marginal case $z=d-1$. 
The mode corresponding to $\beta_-$ is then constant at the
linear level and one has to take into account the non-linear
corrections to lift this degeneracy \cite{Danielsson:2009gi}. 
One finds that due to the non-linearities
the mode is driven to zero logarithmically at $u\to 0$,
\[
\delta h,~\delta j,~ \delta g\propto |\log u|^{-1}\;. 
\]
However, this introduces an additional logarithmic scaling into the function
$f$ on top of the power-law dependence, 
\[
f\propto u^{1-z}\;|\log u|^{-2\frac{d-1}{d-2}}\;,
\]
so that the UV asymptotic is not, strictly speaking,
Lifshitz. Still, it has been argued in \cite{Cheng:2009df} that this
can be interpreted as just a mild violation of the Lifshitz scaling in the
dual theory by the marginally relevant deformation, so that for large
spans in energy the physics is well approximated by the Lifshitz fixed
point.
}. 

\begin{figure}[tb]
\begin{center}
\begin{picture}(470,150)(0,25)
\put(0,0){\includegraphics[scale=0.9]{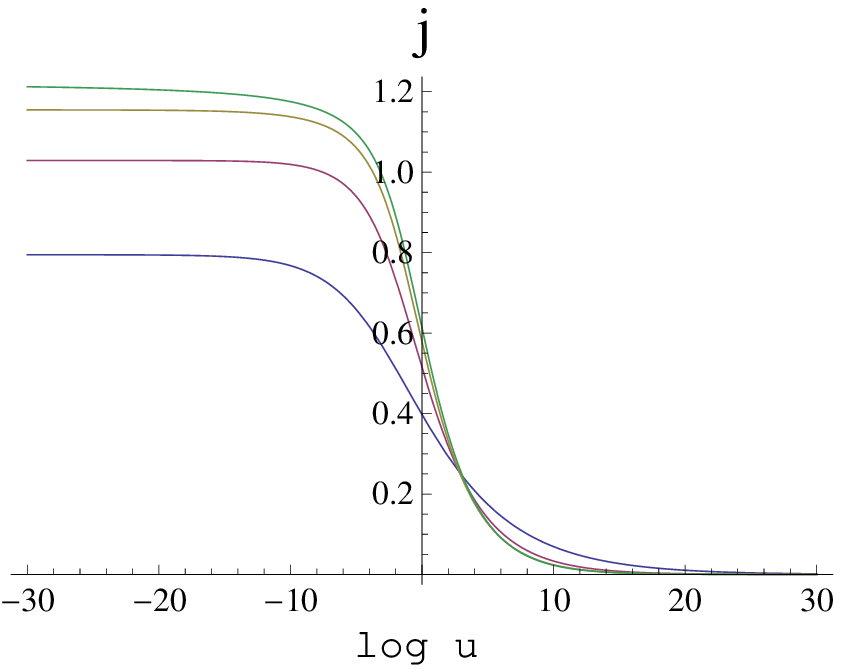}}
\put(250,0){\includegraphics[scale=0.9]{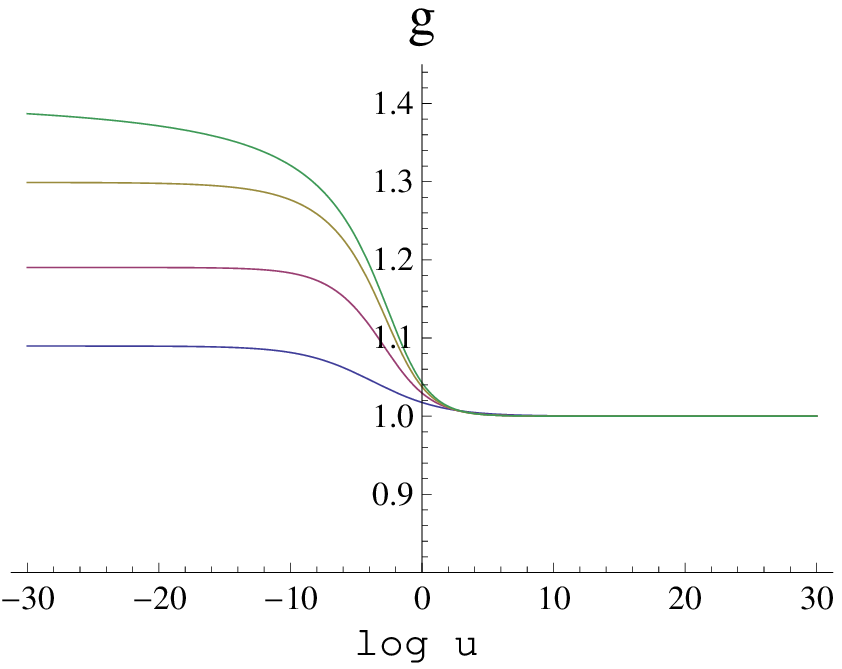}}
\put(10,153){\footnotesize $z=3$}
\put(10,144){\footnotesize $z=2.5$}
\put(10,134){\footnotesize $z=2$}
\put(10,110){\footnotesize $z=1.5$}
\put(260,146){\footnotesize $z=3$}
\put(260,129){\footnotesize $z=2.5$}
\put(260,106){\footnotesize $z=2$}
\put(260,86){\footnotesize $z=1.5$}
\end{picture}
\end{center}
\caption{The vector field and metric profiles for the domain wall
  solutions interpolating between the Lifshitz and AdS spacetimes. The
values of the parameters are $d=4$ and $z=1.5,\; 2,\; 2.5,\; 3$ (from
bottom to top). One may notice from the graphs that the approach to
the UV asymptotics is slower for $z=3$ than for the other
cases, signaling the presence of the marginally relevant mode (see
footnote~\ref{foot:marg}). 
For other values of $d$ the curves are similar.}
\label{Fig:dw}
\end{figure}

This suggests a simple strategy to search for domain wall solutions
with desired properties. Start at large $u$ from AdS slightly
perturbed by the mode $\propto u^{-\al}$ and integrate
Eqs.~(\ref{veceqs}) towards the decreasing value of the coordinate. As
the Lifshitz fixed 
point on the ``$-$'' branch is absolutely stable at $u\to 0$, one may
expect that the solution will be attracted to it at small $u$. This is
indeed confirmed by direct numerical integration of the system
(\ref{veceqs}), the resulting solutions are plotted in
Fig.~\ref{Fig:dw}. It is worth stressing that from the dual viewpoint the
existence of these solutions is highly non-trivial. There one starts
from a fundamental Lifshitz SFT and adds to it the (marginally)
relevant deformation $\tilde O_{\cal A}$. It turns out that the
resulting RG flow is driven to LI in the 
IR\footnote{Interestingly, for the opposite sign of
  the relevant deformation, one obtains a domain wall solution
  connecting to the ``+''-branch of Lifshitz fixed points in the IR
\cite{Braviner:2011kz}.}. 

Let us fix a few notations that will be used in the subsequent
sections. First, from Fig.~\ref{Fig:dw} we see that the transition
from Lifshitz space-time to AdS is quite sharp, which allows to
introduce the notion of the domain wall position (this can be defined,
e.g., as the value of the $u$-coordinate where the function $j(u)$ is
half of its value at $u\to 0$). We will denote this position by
$\Lambda_*^{-1}$. Clearly, there are domain walls with any possible
value of $\Lambda_*^{-1}$ that are related to each other by a
rescaling of the coordinates. From the physical viewpoint, $\Lambda_*$
sets the scale of the transition from non-relativistic to LI
regime. To simplify the formulas we will work
in the units such that 
\be
\label{dmwpos}
\Lambda_*=1\;.
\ee
Second, we will need the coefficients of several leading terms in the
expansion of the domain wall configuration at $u\to\infty$. We write,
\bseq
\label{coeffs}
\begin{align}
&h\approx h_\infty u^{-\al}~,~~~~j\approx j_\infty u^{-\al}\;,\\
\label{deltafgads}
&f\approx 1+f_\infty u^{-2\al}~,~~~~g \approx 1+g_\infty u^{-2\alpha}\;.
\end{align} 
\eseq
In the expressions in the second line we have used that the
perturbations of the metric are induced only at the second order in
$h,j$ as is clear from (\ref{veceq3}), (\ref{veceq4}). The
coefficients in (\ref{coeffs}) are not independent. They can be
determined from (\ref{veceqs}) up to a single constant. However, the
corresponding relations are rather cumbersome and we are not going to
use them in our analysis. The overall normalization of the 
coefficients in (\ref{coeffs}) 
is connected, in turn, to the domain wall position. With
the convention (\ref{dmwpos}) all these coefficients are naturally of
order one. Finally, the numerical solution shows that all functions
$f(u)$,
$g(u)$, $h(u)$, $j(u)$ are monotonically decreasing at all $u>0$. In
particular, this implies that the coefficients $f_\infty$, $g_\infty$,
$h_\infty$, $j_\infty$ are positive. 

A comment is in order. The Lifshitz space-times and the
domain wall solutions described in this section are known to be
geodesically incomplete at $u\to\infty$ where they suffer
from a specific type of singularity 
\cite{Copsey:2010ya,Horowitz:2011gh}.
Despite the fact that 
all
scalar curvature invariants are bounded, 
the tidal forces in a geodesically moving
frame diverge at $u=\infty$ in a finite proper time. 
We will not attempt to provide
any resolution or
physical
interpretation of this singularity\footnote{This issue was discussed
  in \cite{Harrison:2012vy}.}. Instead, we think about the simple setup
based on the Einstein-Proca action (\ref{Sgeom}) as a convenient
toy model illustrating 
general features of 
RG flows with emergent LI, which are expected to be valid in more
sophisticated backgrounds with holographic Lifshitz duals
(e.g. like the supergravity solutions of \cite{Braviner:2011kz}).

\subsection{Probing the flow with the two-point correlator}
\label{sec:correlators}

To study the physical properties of the holographic RG flow we make
use of a probe scalar field $\phi$ that we add to the bulk
action. This implies that the dual strongly coupled theory contains  a
scalar operator $O_\phi$. The coupling of $\phi$ to gravity
is taken to be minimal for simplicity. On the other hand, we introduce
a non-minimal interaction  with the Lorentz breaking vector ${\cal A}_M$
in order to capture the species-dependent effects of LV. This leads 
to the action,
\be
\label{Sphi}
S_\phi=\frac{1}{2b}\int d^{d+1}x\sqrt{-\mathrm{g}}\bigg[\bigg(
-\mathrm{g}^{MN}+
\xi\frac{(ML)^2}{4}{\cal A}^M {\cal
  A}^N\bigg)\d_M\phi\d_N\phi-\mu^2\phi^2\bigg]\;. 
\ee
Due to the $\xi$-coupling the effective metric felt by the field
$\phi$ is different from $g^{MN}$, so that the $\phi$-modes propagate
with a speed different from 1. For $\xi>0$ they are ``subluminal''
(speed smaller than 1), while 
for\footnote{The absolute value of $\xi$ must be small enough for
the combination $1+\xi j^2(u)$ to be positive on the domain wall
solution. Physically, this requirement means that the term with
time-derivatives in the action (\ref{Sphi}) is
everywhere positive, so that the spectrum of excitations does not
contain ghosts.} $\xi<0$ they are
``superluminal'' (speed greater than 1). 
Note that we could add several probe fields with different values of the
parameter $\xi$ meaning that this parameter encodes the
species-dependent information. 

Using the operator $O_\phi$ one can construct various observables
probing the properties of the RG flow dual to the domain wall solution.
In this section we concentrate on the
two-point function,
\be
\label{corr0}
{\cal G}_\phi(\omega,{\bf k})=\langle O_\phi(\omega, {\bf k}) 
O_\phi(-\omega,-{\bf k})\rangle\;. 
\ee
At small distances, where the relevant perturbation of the Lifshitz
fixed point is not 
effective, one expects this correlator to obey the Lifshitz scaling. On the other
hand, at large distances, i.e. at small energies
and momenta, its behavior must be dominated by the LI infrared fixed
point and therefore it must exhibit approximate LI. 
Our aim will be to verify the latter assertion.
Note that it is not immediately obvious, given that 
the holographic
prescription for the calculation of the correlator involves taking
the limit of certain quantities at $u\to 0$, i.e. deep inside the
Lifshitz region. Additionally, we will find the form of the LV
corrections to the correlator at low energies.

The calculation is based on the solutions of the equation of motion
for the scalar field obtained from (\ref{Sphi}). This reads in the
Fourier representation,
\be
\label{scalareq}
\phi''+\bigg(\frac{f'}{f}-\frac{g'}{g}-\frac{d-1}{u}\bigg)\phi'
-g^2\bigg(\frac{1+\xi j^2}{f^2}w^2+k^2+\frac{(\m L)^2}{u^2}\bigg)\phi=0\;,
\ee
where primes denotes the derivatives with respect to $u$ and we performed
the Wick rotation to the  
Euclidean frequency, $w=-i\omega$. This equation cannot be solved
exactly because we do not have analytic expressions for the functions $f,g,j$
of
the domain wall background. However, for our purposes the explicit
form of the $\phi$-modes
will not be required. It
suffices to know their qualitative features, which we now discuss. 

\begin{figure}[tb]
\centerline{\includegraphics[scale=0.5]{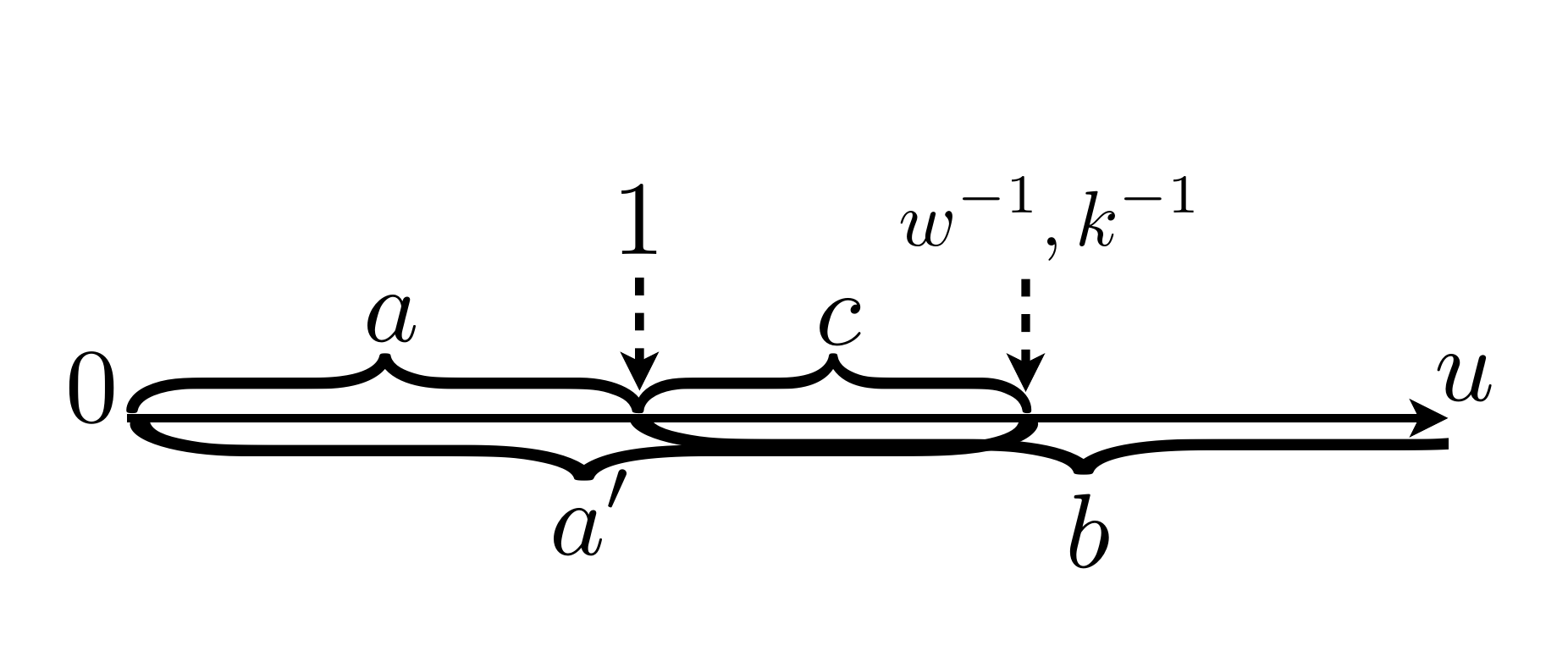}}
\caption{The regions on the $u$-axis that enter into the construction
  of the solutions to Eq.~(\ref{scalareq}). The case of small
  frequency and momentum is shown.}
\label{Fig:regions}
\end{figure}

In the Lifshitz region $u\ll 1$ the background functions can be
replaced by their asymptotic form\footnote{We assume that $z$ is
  strictly smaller than $d-1$ to avoid the
  complications appearing in the case of marginally relevant
  deformation.} 
(\ref{fz}),
(\ref{fixedps}), where the ``-'' branch
 must be 
taken\footnote{The expressions in terms of the
critical exponent are, 
\[
g_-=\sqrt{\frac{z^2+(d-2)z+(d-1)^2}{d(d-1)}}~,~~~~
j_-=\sqrt{\frac{d(d-1)^2(z-1)}{z^2+(d-2)z+(d-1)^2}}\;.
\]}.
Then one can choose two linearly independent solutions of
(\ref{scalareq}) that have the form of a simple power-law at $u\to 0$,
\bseq
\label{phias}
\begin{align}
&\phi_{a,\pm}= u^{\lambda_{\pm}}~,~~~u\to 0\;,\\
\label{lambdas}
&\lambda_\pm=\frac{d+z-1}{2}\pm
\sqrt{\bigg(\frac{d+z-1}{2}\bigg)^2+(g_-\mu L)^2}\;.
\end{align}
\eseq
Here the subscript ``$a$'' reflects the fact that these solutions are
defined by their behavior 
in the Lifshitz part of the space-time, to which we will refer
as the ``$a$-region'' for short, see
Fig.~\ref{Fig:regions}. 

We need a linear combination of the
solutions (\ref{phias}) that decays at infinity, $u\to\infty$. This
will be denoted 
by $\phi_{b}(u)$, being defined by the behavior in the
``$b$-region'' $u\gg 1$. We write,
\be
\label{phiba}
\phi_{b}(u)=T_{-}\phi_{a,-}(u)+T_{+}\phi_{a,+}(u)\;,
\ee
where the coefficients $T_{\mp}$ depend on the frequency $w$ and
momentum $k$. According to the dictionary of the holographic correspondence
the amplitude $T_{-}$ is identified
with the source for the operator $O_{\phi}$. The correlator
(\ref{corr0}) is then
expressed through the variation of 
the classical action (\ref{Sphi}) evaluated on the
solution (\ref{phiba}),
\[
\begin{split}
{\cal G}_\phi(w,{\bf k})&\propto
\frac{\delta^2 S_{\phi}}{\delta T_{-}(w,{\bf k})\delta T_{-}({-w,-{\bf k}})}
+\text{c.t.}\\
&\propto\frac{1}{T_-^2}\lim_{u\to 0}\bigg(u^{1-d}\,\frac{f}{g}\phi_{b}\phi_{b}'
+\text{c.t.}\bigg)\;.
\end{split}
\]
The proportionality signs here emphasize that we do not keep track of
the overall normalization of the correlator focusing only on
its frequency and momentum dependence. 
The notation ``c.t.'' stands for the counterterms that 
must be included to remove the divergencies
that arise in taking the limit indicated in the second line. 
Using (\ref{phiba}) we obtain,
\be 
\label{corr11}
{\cal G}_\phi(w,{\bf k})\propto \lim_{u\to 0} \bigg(
u^{1-d}\,\frac{f}{g}\phi_{a,-}\phi_{a,-}'+\text{c.t.}\bigg)
+\frac{(\lambda_++\lambda_-)f_0}{g_-}\cdot
\frac{T_+(w,k)}{T_-(w,k)}\;.
\ee
Only the first term diverges at $u\to 0$ and must be
cancelled by the counterterms. 
Importantly, this term is polynomial
in $w$ and $k$, meaning that it is local in space and time. Indeed,
at $u\to 0$ the contributions containing $w,k$ in Eq.~(\ref{scalareq}) are
subleading and the solution $\phi_{a,-}$ can be obtained by iterating
them in 
a perturbative expansion. As a result 
$\phi_{a,-}$ is obtained in the form of a power series in $w^2$ and
$k^2$. Only a finite number 
of terms in the series lead to divergencies when substituted into
(\ref{corr11}), implying the above assertion.   
Thus we conclude that the first term in (\ref{corr11}) does not affect 
the correlator of
operators taken at different spacetime points, which is given by the
non-analytic part of ${\cal G}_\phi(w,k)$. In what follows we will 
discard polynomial contributions into the correlator. Thus, we
are left with the second term in (\ref{corr11}). In other words, the
correlator is given by the ratio of the coefficients in the linear
relation (\ref{phiba}). The situation is exactly the same as in the standard
case of the relativistic AdS/CFT correspondence, cf. \cite{Freedman:1998tz}. 

Before concentrating on the case of small frequencies and
momenta, which is of the main interest to us, we briefly discuss
the opposite regime $w,k\gg
1$. The decaying solution $\phi_{b}$ is predominantly localized
at $u\lesssim 1/k$ and thus for large momenta it is entirely controlled
by the Lifshitz region $u\ll 1$. By an appropriate rescaling of
variables one finds that it has the form,
\[
\phi_{b}=\tilde\phi_{b}\bigg(g_-ku\,;\,\frac{1+\xi j_-^2}{f_0^2g_-^{2z-2}}
\,\frac{w^2}{k^{2z}}\,,\,g_-\mu L\bigg)\;,
\]
where $\tilde\phi_b(x;\eta_1,\eta_2)$ is the decaying solution of the equation
\[
\tilde\phi''+\frac{2-d-z}{x}\,\tilde\phi'-
\bigg(\eta_1\, x^{2z-2}+1+\frac{\eta_2^2}{x^2}\bigg)\tilde\phi=0\;.
\]
Substitution into (\ref{phiba}), (\ref{corr11}) yields,
\[
{\cal G}_{\phi}\propto k^{\lambda_+-\lambda_-}\,
\tilde {\cal G}\bigg(\frac{1+\xi j_-^2}{f_0^2g_-^{2z-2}}
\,\frac{w^2}{k^{2z}}\,,\,g_-\mu L\bigg)\;.
\]
For the case $d=3$, $z=2$ the function $\tilde{\cal G}$ was found
explicitly in \cite{Kachru:2008yh}.
We see that at large momenta the correlator obeys the Lifshitz scaling
with the critical exponent $z$. The overall scaling of the
correlator with momentum implies that the 
dimension of the operator
$O_\phi$ in the Lifshitz fixed point is $\lambda_+$ (recall that 
$\lambda_+-\lambda_-=2\lambda_+-z-d+1$), which is consistent with the
holographic interpretation of the solution $\phi_{a,+}$ (see
(\ref{phias})) as the expectation value of $O_\phi$.
Note also that the dependence of the
correlator on frequency and momentum is not
universal: it contains the species-dependent coupling $\xi$.

We now study in detail the behavior of ${\cal G}_\phi$ in the
low-energy regime $w,k\ll 1$. In this case the structure of the
solution $\phi_b$ is more complicated: it feels all of the domain wall
configuration and has a tail extending deep inside the $b$-region. In
the latter region the space-time is approximately AdS, so one can find
$\phi_b$ analytically by expanding in the small deviations of the
background functions $f,g,h$ from their AdS form. On the other hand,
at $u\ll w^{-1},k^{-1}$ (we will call this ``$a'$-region'') the terms
proportional to the frequency and momentum in Eq.~(\ref{scalareq}) can
be treated as perturbations and the solution can be expanded 
as a
power-series in $w^2$, $k^2$. This suggests the following perturbative
strategy to construct $\phi_b$: find the two expansions above to a
given order in the corresponding domains and match them in the overlap 
$1\ll u\ll w^{-1},k^{-1}$. The latter overlap will be referred to 
as the ``$c$-region''. The regions involved in the analysis are
depicted in Fig.~{\ref{Fig:regions}}.

Consider first the leading order. In the $b$-region one sets $f=g=1$,
$j=0$ and obtains,
\be
\label{phib-0}
\phi_{b}^{(0)}=u^{d/2} K_{\nu}(p_Eu)\;,
\ee 
where we use the same notations as in Sec.~\ref{ssec:propel} and $\nu$
is defined in (\ref{nu}). 
In the $c$-region this function is matched 
to the linear combination,
\be
\label{phib-dec0}
\phi_{b}^{(0)}(u)=U_{-}^{(0)}\,\phi_{c,-}^{(0)}(u)+
U_{+}^{(0)}\,\phi_{c,+}^{(0)}(u)\;,
\ee
where $\phi_{c,\pm}^{(0)}$ are the solutions of
Eq.~(\ref{scalareq}) with $w$ and $k$ set to zero. They are fixed by
their asymptotics at large $u$,
\be
\label{phic0}
\phi_{c,\pm}^{(0)}\approx u^{d/2\pm\nu}~~~~\text{at}~u\gg 1\;. 
\ee 
Using the expansion of the Macdonald function at small values of the
argument yields,
\be
\label{U0s}
U_{\pm}^{(0)}=2^{\mp \nu-1}\Gamma(\mp\nu)p_E^{\pm\nu}\;.
\ee
To find the coefficients $T_{\pm}$ entering into (\ref{phiba}) we need
to rotate from the basis $\phi_{c,\pm}^{(0)}(u)$ in the space of
solutions to Eq.~(\ref{scalareq}) with $w=k=0$ to the basis
$\phi_{a,\pm}^{(0)}$ defined by the asymptotics~(\ref{phias}). We write 
\be
\label{phica0}
\phi_{c,-}^{(0)}=V_{--}^{(0)}\phi_{a,-}^{(0)}+V_{-+}^{(0)}\phi_{a,+}^{(0)}
~,~~~~
\phi_{c,+}^{(0)}=V_{+-}^{(0)}\phi_{a,-}^{(0)}+V_{++}^{(0)}\phi_{a,+}^{(0)}\;.
\ee 
The coefficients $V_{\pm,\pm}^{(0)}$ are sensitive to the shape of the
domain wall background and cannot be found explicitly. For our
purposes it is sufficient to know 
that they do not contain any dependence
on $w$ or $k$. Combining the linear relations (\ref{phica0}),
(\ref{phib-dec0}) and evaluating the ratio of the resulting
coefficients $T_\pm$ we obtain,
\be
\label{corr4}
{\cal
  G}_\phi(w,k)\propto 2^{-2\nu}\frac{\Gamma(-\nu)}
{\Gamma(\nu)}\frac{\big|V^{(0)}\big|}{\big(V_{--}^{(0)}\big)^2}
\cdot p_E^{2\nu}+\ldots\;,
\ee
where $\big|V^{(0)}\big|$ is the determinant of the 
$2\times 2$-matrix composed of the coefficients\footnote{One can show
  that $\big|V^{(0)}\big|>0$, see Appendix~\ref{app:corr}. This ensures
  that the correlator has correct analytic properties: its imaginary
  part is positive on the upper edge of the cut at 
$p^2\equiv -p_E^2>0$.}
$V^{(0)}_{\pm,\pm}$ and dots stand for the terms with higher powers of
$p_E^{2\nu}$ that are suppressed at $p_E\ll 1$. 
In deriving (\ref{corr4}) we have subtracted a constant (frequency and
momentum independent) contribution. The expression (\ref{corr4}) is
manifestly Lorentz invariant. It has the right dependence on the
momentum for the two-point function of the scalar operator with dimension
(\ref{dimO}), which is indeed the dimension of $O_\phi$ at the
relativistic fixed point. Thus we have checked that at large distances
the behavior of the correlator ${\cal G}_\phi$ changes from Lifshitz
to LI, as expected.

Next we want to investigate the LV corrections to (\ref{corr4}). 
This requires going to the next order in the perturbative scheme for
the solution $\phi_b$.
The calculation is rather tedious and we
present it in Appendix~\ref{app:corr}. The final result has the 
form\footnote{Recall
that $\alpha$ has been defined in Sec.~\ref{sec:geometry} and
is related to the asymptotics of the domain wall background, see
(\ref{coeffs}).},
\be
\label{corr5}
{\cal
  G}_\phi(w,k)\propto 2^{-2\nu}\frac{\Gamma(-\nu)}
{\Gamma(\nu)}\frac{\big|V^{(0)}\big|}{\big(V_{--}^{(0)}\big)^2}
\cdot p_E^{2\nu}\cdot\big[1+w^2\big(C_1\,p_E^{-2+2\alpha}+C_2+
C_3p^{2\n}_E\big)
+\ldots\big]\;,
\ee
where the coefficients
$C_{1,2,3}$ are given in (\ref{C123}) and
dots stand for terms that are either of higher power in frequency and
momentum, 
or are LI, i.e. depend on $w,k$ only through
the combination $p_E$. Note that we have kept 
the $C_3$-term in spite of the fact that it is subleading compared to
the $C_2$-contribution; the reason for this will become clear
shortly. One observes that the LV corrections are in general different
for different species: the formulas (\ref{C123}) for the coefficients
depend both on the parameter $\n$ related to the dimension of the
operator 
$O_\phi$ and on the species-specific coupling $\xi$.

As long as $\al<1$ the $C_1$-term in (\ref{corr5}) dominates meaning
that the scaling of the LV correction with energy is
determined by the dimension of the vector operator $O_{\cal A}^\mu$,
see Eq.~(\ref{dimOA}). If $\al>1$ the $C_2$-term becomes dominant and
the LV correction scales as $w^2$. Finally, for $\al>1+\n$ the
$C_1$-term is small compared to the $C_3$-contribution and the leading
corrections are completely controlled by the properties of the operator
$O_\phi$ itself. Note, however, that due to the upper bound on $\al$
discussed in Sec.~\ref{sec:geometry} this situation cannot be realized
within the considered setup for spacetimes of low dimension.

The structure of the corrections (\ref{corr5}) admits nice holographic
interpretation. They correspond to an irrelevant perturbation of the
relativistic CFT emerging at the IR fixed point by the leading LV
operators. Apart from the operator $O_{\cal A}^t$ discussed in
Sec.~\ref{sec:geometry}, we have at our disposal the time derivative
of $O_\phi$. Omitting dimensionless coefficients we write 
schematically for the perturbation Lagrangian,
\be
\label{dLCFT}
\delta{\cal L}_{CFT}=\frac{O_{\cal A}^t}{\Lambda^{\al}_*}
+\frac{{\dot O}^2_\phi}{\Lambda_*^{2+2\n}}
+\bigg(O_\phi+\frac{\dot O_\phi}{\Lambda_*}\bigg)\cdot J+\ldots\;,
\ee 
where we have restored the LV scale $\Lambda_*$ in appropriate
powers. Dots stand for operators of higher dimension that are further
suppressed by $\Lambda_*$. Note that we have included the interaction
of $O_\phi$ with the source $J$ that also receives LV
contributions due to the RG flow. Now, various terms in (\ref{corr5})
are identified with the contribution of (\ref{dLCFT}) into the
perturbative calculation of the $O_\phi$-propagator around the IR
fixed point. The $C_1$-term comes from the double insertion\footnote{A
single insertion gives zero due to vanishing of the three-point
function, $\langle O_\phi O_\phi O_{\cal A}^\m\rangle=0$, in the
unperturbed CFT. The latter property follows from the invariance of
the bulk theory under the flip of sign ${\cal
  A}_M\mapsto {\cal A}_M$.} of the
first term of (\ref{dLCFT}) into the propagator, the $C_3$-term is
identified with the insertion of the second term, and the $C_2$-term
comes from the modification of the coupling to the source.
An immediate consequence of this identification is that the LV
correction of the $C_2$-type is not universal in the sense that it
will not appear in the observables which do not involve external
sources. Indeed, we will see in the next subsection that this type of
corrections does not show up in the dispersion relations of the bound
state present in a confining version of the theory. On the contrary,
the LV corrections of the $C_1,C_3$-types are expected to be universal
as they are intrinsic to the theory itself.

\subsection{Bound states}
\label{sec:states}

To make contact with particle physics, the model of the previous
subsection must be supplemented by a mechanism providing a discrete
particle spectrum. This is achieved by cutting the space-time with a
brane located at $u=\Lambda^{-1}$ and keeping only the portion
$0<u<\Lambda^{-1}$. Note that in this in way one shields the curvature
singularity at $u\to\infty$ mentioned at the end of
Sec.~\ref{sec:geometry}. In Appendix~\ref{app:IRbrane} we show that
the brane energy-momentum tensor required for consistent matching with
the domain wall solution is a sum of a constant negative tension plus
a contribution that satisfies the null energy condition and thus
can be provided by some regular matter. This setup gives
rise to a discrete spectrum of KK modes of the scalar field $\phi$ that
by the holographic correspondence are associated to bound states in
the dual theory. From the dual viewpoint the bound states appear
due to the introduction of an IR cutoff (confining scale) $\Lambda$
set by the brane, to which we will refer as ``IR brane'' in what
follows. We will assume the confining scale to be much lower than the
scale of LV, $\Lambda\ll \Lambda_*$, which under the convention (\ref{dmwpos}) is
equivalent to
\[
\Lambda\ll 1\;.
\]  
Our aim is to determine the dispersion relations of the bound states
(equivalently, KK modes). Similarly to the model of Sec.~\ref{sec:RS}, 
 we introduce a
mass $\m_{IR}$ for the field $\phi$ on the IR
brane that will be tuned to make the lightest
KK mode gapless. 

It is convenient to put the eigenmode equation (Eq.~(\ref{scalareq})
with $w^2$ replaced by $-\omega^2$) into the form of a quantum
mechanical eigenvalue problem,
\be
\label{eqeigen}
\big(\hat H_{(0)}+V(u)\big)\phi=(\omega^2-k^2)\phi\;,
\ee 
where the operator $\hat H_{(0)}$ is defined as,
\[
\hat H_{(0)}=\frac{f^2}{g^2(1+\xi j^2)}\bigg[-\frac{d^2}{du^2}
-\bigg(\frac{f'}{f}-\frac{g'}{g}-\frac{d-1}{u}\bigg)\frac{d}{du}
+g^2\frac{(\m L)^2}{u^2}\bigg]\;,
\]
and
\be
\label{H1}
V(u)=\bigg[\frac{f^2}{1+\xi j^2}-1\bigg]\; k^2\;.
\ee
Equation (\ref{eqeigen}) is supplemented by the boundary conditions,
\bseq
\label{bcc}
\begin{align}
\label{bcc1}
&\phi\to 0~~~~ \text{at}~~ u\to 0\;,\\
\label{bcc2}
&\frac{\phi'}{b}+gL\Lambda\bar\m_{IR}^2\phi\bigg|_{u=\Lambda^{-1}}=0\;.
\end{align}
\eseq
Note that we impose the Dirichlet condition in the UV. The IR
boundary condition (\ref{bcc2}) is due to the mass
term on the brane, cf. (\ref{AdSbc2}). It is straightforward to check
that the operator $\hat H_{(0)}$ is Hermitian
in the space of functions satisfying these boundary conditions
with the scalar product
\be
\label{product}
\bra{\phi_1}\phi_2\rangle=
\int_0^{\Lambda^{-1}} du\, u^{1-d}\;\frac{g(1+\xi j^2)}{f}\;\phi_1^*(u)\phi_2(u)\;.
\ee
The key idea of the calculation 
below is to treat the potential (\ref{H1})
as a small 
perturbation. This is certainly true for low-lying KK modes with small
enough momenta. Indeed, $V(u)$ is localized at
$u\lesssim 1$, while the low-lying modes are expected to spread over 
the whole interval $u< \Lambda^{-1}$. 
Another crucial observation is that the
leading operator $\hat H_{(0)}$ does not contain any dependence on the
spatial momentum $k$. Thus neglecting $V$ we obtain
in the leading order a set of LI dispersion relation 
\[
\omega_n^2=k^2+m_n^2\;,
\]
where the masses $m_n^2$, $n=0,1,2,\ldots$, 
are given by the eigenvalues of $\hat H_{(0)}$. The LV
corrections are then found using the standard techniques of 
quantum-mechanical perturbation theory.

\subsubsection{Unperturbed eigenfunctions}

To proceed along these lines we will need the matrix elements of
$V(u)$ between the eigenfunctions of the unperturbed Hamiltonian $\hat
H_{(0)}$. Thus, our first task is to find these eigenfunctions. We focus
on the light KK modes with the masses of order the confinement scale
$\Lambda$. This implies $m_n\ll 1$, and we can use the same technique
as in Sec.~\ref{sec:correlators}: find the solutions separately in the
$a'$-region $u\ll 1/m_n$, in the $b$-region $u\gg 1$ and match them
in
the overlap 
$1\ll u\ll 1/m_n$ ($c$-region). In the $a'$-region 
the mass can be neglected altogether and one is left with
the equation $\hat H_{(0)}\phi=0$. This equation already appeared in
Sec.~\ref{sec:correlators} and its solution vanishing at 
$u\to 0$ has been
denoted by $\phi_{a,+}^{(0)}$. 
Thus we
write,
\be
\label{phina}
\phi_n(u)=a_n \phi_{a,+}^{(0)}(u)~,~~~~u\ll 1/m_n\;,
\ee
where the normalization coefficient $a_n$ will be fixed below.
On the
other hand, in the $b$-region the functions 
$f,g,j$ can be set to their asymptotic values, $f=g=1$, $j=0$, and the
solution reads,
\be
\label{phinb}
\phi_n(u)=b_n u^{d/2} J_\n(m_n u)+\tilde b_n u^{d/2}Y_\n(m_n u)
~,~~~~u\gg 1\;.
\ee 
Using the expansion of the Bessel functions in the $c$-region and the
asymptotics,
\be
\label{phiaasym}
\phi_{a,+}^{(0)}\approx
\frac{V_{--}^{(0)}}{\big|V^{(0)}\big|}u^{d/2+\n}
-\frac{V_{+-}^{(0)}}{\big|V^{(0)}\big|}
u^{d/2-\n}~,~~~~u\gg 1\;,
\ee
obtained by inverting Eqs.~(\ref{phica0}), we find the relations,
\be
\label{anbn}
b_n=a_n\, \frac{V_{--}^{(0)}}{\big|V^{(0)}\big|}\, 2^\n\Gamma(1+\n)
\,m_n^{-\n}~,~~~~
\tilde b_n=a_n\, \frac{V_{+-}^{(0)}}{\big|V^{(0)}\big|}\,
2^{-\n}\sin{\pi\n}\Gamma(1-\n)
\,m_n^{\n}\;.
\ee
One observes that the coefficient $\tilde b_n$ is suppressed by
$m_n^{2\n}\ll 1$ compared to $b_n$. Thus, the second term in
(\ref{phinb}) can be neglected when $m_nu\gtrsim 1$. As the KK masses
are expected to be of order\footnote{Except, probably, the zero mode
  that must be treated separately, see below.} 
$\Lambda$, this implies that the second
term in (\ref{phinb}) is irrelevant over most of the interval
$u<\Lambda^{-1}$. Taking this into account we obtain from the boundary
condition (\ref{bcc2}) the eigenvalue equation, 
\be
\label{bccc}
J_{\n+1}(m_n/\Lambda)-\frac{\Lambda}{m_n}
\bigg(bL{\bar\m}_{IR}^2+\frac{d}{2}+\n\bigg)J_{\n}(m_n/\Lambda)=0\;.
\ee

At this stage, to somewhat simplify the subsequent formulas, we will
impose the requirement that $\hat H_{(0)}$ has a massless mode. This 
has the form (\ref{phina}) in the whole interval. Using the asymptotic
expression (\ref{phiaasym}), where we neglect the second term, we find
that it satisfies the boundary condition (\ref{bcc2}) if 
\[
{\bar\m}_{IR}^2=-\bigg(\frac{d}{2}+\n\bigg)\frac{1}{bL}\;.
\] 
Note that according to Eq.~(\ref{bccc}) 
the masses of other modes are then given simply by the
zeros of $J_{\n+1}$.

It remains to fix the 
normalization of the modes. Using the scalar product (\ref{product})
we impose 
\[
\int_0^{\Lambda^{-1}} du\,u^{1-d}\,\frac{g(1+\xi
  j^2)}{f}\big(\phi_n(u)\big)^2=1\;. 
\]
The
integral is saturated at $u\gg 1$, so we use Eqs.~(\ref{phinb}) and
(\ref{phiaasym}) for the massive and massless modes respectively
(where in both expressions we neglect the second term). This yields,
\be
\label{bn1}
a_0=\sqrt{2(1+\n)}\,\Lambda^{1+\n}\frac{\big|V^{(0)}\big|}{V_{--}^{(0)}}~,~~~~
b_n=\sqrt 2 \Lambda |J_{\n}(m_n/\Lambda)|^{-1}~,~~~~n\geq 1\;.
\ee
In deriving the second expression we have used that $J_{\n+1}(m_n/\Lambda)=0$.
Equations (\ref{phina}) --- 
(\ref{anbn}), (\ref{bn1}) 
completely specify the eigenmodes of $\hat H_{(0)}$.

\subsubsection{First-order LV corrections}

We are now ready to evaluate the LV corrections to the dispersion
relations of the modes. The first-order correction has the form,
\be
\label{domega1n}
\delta\omega_{n,(1)}^2=\bra{\phi_n}V\ket{\phi_n}=
k^2\int_0^{\Lambda^{-1}}du\,u^{1-d}\,
g\bigg(f-\frac{1+\xi j^2}{f}\bigg)(\phi_n)^2\;.
\ee
The two factors in the integrand 
are peaked at different values of
$u$. The combination in the brackets deviates 
from zero at $u\lesssim
1$, with a tail that extends to larger $u$ falling down as 
$u^{-2\alpha}$ (see (\ref{coeffs})). On the other hand, the
square of the wavefunction has maximum at $u\gtrsim 1/m_n\gg 1$ and
decreases as $u^{d+2\n}$ towards smaller $u$. Depending on which
factor dominates, the integral is
saturated either in the IR (i.e. $u\gg 1$) or in the UV ($u\lesssim
1$). Let us consider the corresponding options case by case.

{\bf (i)} If $\alpha<1+\n$ the integral is saturated in the IR\footnote{As
  mentioned in Sec.~\ref{sec:geometry}, 
this case is actually always
  realized in the model at hand, unless $d\geq 7$.}. In this
region we use the asymptotics (\ref{coeffs}) and the expression
(\ref{phinb}) for the wavefunction. 
Expanding the integrand to the leading order in $u^{-2\alpha}$ we obtain,
\bseq
\label{deltaomega1}
\begin{align}
\label{domega1n4}
&\delta\omega_{0,(1)}^2=k^2 (2f_\infty-\xi j_\infty^2)
\frac{1+\n}{1+\n-\alpha}\; \Lambda^{2\alpha}\;,\\
\label{domega1n1}
&\delta\omega_{n,(1)}^2=k^2 (2f_\infty-\xi j_\infty^2) 
\frac{2\Lambda^2 m_n^{-2+2\alpha}}{\big(J_\n(m_n/\Lambda)\big)^2}
\int_0^{m_n/\Lambda}dx\,x^{1-2\alpha}\big(J_\n(x)\big)^2~,~~~n\geq 1\;.
\end{align}
\eseq
The combination multiplying $k^2$
on the r.h.s. has the physical meaning of (twice) the correction to the
propagation velocity of the mode.
Notice that it
depends on the non-minimal coupling $\xi$ of the scalar field to the
vector, as well as on the mass of the mode, implying that it is 
non-universal. At the same time it is clearly suppressed by the RG
flow. Indeed, assuming $m_n\sim\Lambda$ and restoring explicitly the
LV scale $\Lambda_*$, we see that all
corrections to the velocity are of order
$(\Lambda/\Lambda_*)^{2\alpha}$.

One may wonder if the suppression persists for heavier modes. To
answer this question let us consider the limit of
Eq.~(\ref{domega1n1}) for $m_n\gg\Lambda$ (but still assuming $m_n\ll
1$). Depending on the value of $\al$ the integral entering this
expression converges or not when the upper limit tends to
infinity; this gives two cases,
\be
\label{domega1n2}
\delta\omega_{n,(1)}^2\propto 
k^2(2f_\infty-\xi j_\infty^2)\times
\begin{cases}
\Lambda^{2\al}~,&\al<1/2\\
\Lambda m_n^{-1+2\al}~,&\al>1/2
\end{cases}
\ee
where we have omitted irrelevant numerical factors of order one. We
see that in the first case the correction to the mode velocity does
not scale with the mass, while in the second case it grows. However,
in both cases the corrections are small as long as $m_n$ does not
exceed 1, where the formula (\ref{domega1n2}) stops being applicable
anyway. 

{\bf (ii)}
For $\al>1+\n$ the integral in (\ref{domega1n})
is saturated in the UV, $u\lesssim 1$. Substituting the wavefunction
in the form (\ref{phina}) and using (\ref{anbn}), (\ref{bn1}) one obtains,
\bseq
\label{deltaomega11}
\begin{align}
\label{domega1n8}
&\delta\omega_{0,(1)}^2=k^2\cdot 2(\n+1)B\,\Lambda^{2+2\n}\;,\\
\label{domega1n5}
&\delta\omega_{n,(1)}^2=k^2\frac{2^{1-2\n}B}{(\Gamma(\n+1))^2}
\frac{\Lambda^2m_n^{2\n}}{\big(J_\n(m_n/\Lambda)\big)^2}\;,~~~~
n\geq 1\;,
\end{align}
\eseq
where
\be
\label{calI}
B=\bigg(\frac{\big|V^{(0)}\big|}{V^{(0)}_{--}}\bigg)^2
\int_0^\infty du\, u^{1-d}\,g\bigg(f-\frac{1+\xi
  j^2}{f}\bigg) \big(\phi_{a,+}^{(0)}\big)^2\;.
\ee
Note that we have exploited the fast falloff of the integrand at $u>1$
and extended the integration to infinity. 
One observes that for $m_n\sim\Lambda$ the corrections to the
velocities of the modes are suppressed by
$(\Lambda/\Lambda_*)^{2+2\n}$. For larger masses the corrections
increase and at $m_n\gg\Lambda$ we obtain
\[
\delta\omega_{n,(1)}^2\propto k^2 B\,\Lambda m_n^{-1+2\n}\;.
\]
Still this is parametrically small for all $m_n\ll 1$.

The structure of the LV corrections to the dispersion relations 
found above fits naturally into the holographic interpretation of the
model. The expressions (\ref{deltaomega1}), (\ref{deltaomega11}) are
interpreted respectively as the contributions of the first and second
operator in the Lagrangian (\ref{dLCFT}) describing the perturbation
from the LI fixed point.

\subsubsection{Higher-order corrections}

In principle, the quantum-mechanical perturbation theory allows to
compute higher-order corrections to the dispersion relations of the KK
modes. The $l$th correction will involve sums over products of $l$
matrix elements of the LV potential $V(u)$, see (\ref{H1}). As the
latter is proportional to $k^2$, one will obtain in this way the
dispersion relation in the form (\ref{disprel2}), as a series in
powers of the spatial momentum. This is precisely the form used in the
literature to analyze the phenomenological constraints on LV. However,
the complexity of the formulas increases rapidly as one goes to higher
orders $l$, making calculation of the coefficients in this series impractical
beyond the case $l=2$, which we are now going to consider. 
Not to overload the paper, we restrict
the analysis to the zero mode. Apart from giving the scale $M_4$ that
suppresses the $k^4$ term in the dispersion relation, this analysis
will provide information about the domain of validity of   
the
perturbative calculation.

The starting formula for the second-order correction to the frequency
of the zero mode is 
\be
\label{PTformulas}
\delta\omega_{0,(2)}^2=-\sum_{n=1}^{\infty} m_n^{-2}\;
|\bra{\phi_0}V\ket{\phi_n}|^2\;.
\ee 
Note that this is negative implying that the $k^4$ contribution comes
with the minus sign. This appears to be a general feature of the
holographic setup valid beyond the specific model of this
paper. It is a direct consequence of the fact that the second-order
correction to the frequency of the lowest KK mode is always negative.

The evaluation of the matrix element appearing in (\ref{PTformulas})
proceeds differently depending on which of the quantities, $\al$ or
$1+\n$, is larger. We have already encountered this situation above in
the computation of the first-order LV contribution. Thus we have:

Case {\bf (i)}: $\al<1+\n$. The integral defining the matrix element
is saturated at $u\gg 1$, which yields,
\be
\label{matrel}
\bra{\phi_0} V\ket{\phi_n}=k^2(2f_\infty-\xi j_\infty^2)
\frac{2\sqrt{1+\n}\,\Lambda^{2+\n}m_n^{-2-\n+2\al}}{|J_\n(m_n/\Lambda)|}
\int_0^{m_n/\Lambda}dx\; x^{1+\n-2\al}J_\n(x)\;.
\ee  
We need the asymptotics of this expression at $n\gg 1$. There are again
two possibilities depending on whether or not the remaining integral
converges at $m_n/\Lambda\to\infty$. Using the formula
\[
J_\n(x)\approx \sqrt{\frac{2}{\pi x}}
\cos\bigg(x-\frac{\pi\n}{2}-\frac{\pi}{4}\bigg)
~~~~\text{at}~~x\to \infty
\]
and integrating by parts we obtain the leading behavior,
\[
\int_0^{m_n/\Lambda}dx\;x^{1+\n-2\al}J_\n(x)=\const_1+\const_2\cdot 
(m_n/\Lambda)^{-\frac{1}{2}+\n-2\al}\;.
\]
In this derivation we have used that the masses are given by the zeros
of $J_{\n+1}$, hence at $n\gg 1$,
\[
m_n\approx\bigg(\frac{\pi\n}{2}+\frac{\pi}{4}+\pi n\bigg)\Lambda\;.
\]
In this way we obtain the following expression for the matrix element
at large $n$,
\[
\bra{\phi_0} V\ket{\phi_n}\propto k^2(2f_\infty-\xi j_\infty^2) 
\Lambda^{2\al}
\times\begin{cases}
n^{-2}\;,&\al<-1/4+\n/2\\
n^{-\frac{3}{2}-\n+2\al}\;,&\al>-1/4+\n/2\;.
\end{cases}
\]

Substituting this into (\ref{PTformulas}) we observe that the sum over
$n$ converges provided $\al<1+\n/2$. To understand the origin of 
the divergence that
appears in the opposite case, one recalls that Eq.~(\ref{matrel}) is
valid only for masses smaller than the LV scale, $m_n\ll 1$. 
The divergence is an artifact of this low-mass approximation: the
total correction to the energy level is of course finite in quantum
mechanics. For a crude estimate we can just cut off the sum at
$n_{max}\sim \Lambda^{-1}$. Combining everything together we arrive
at,
\be
\label{domega21}
\delta\omega_{0,(2)}^2\propto -k^4(2f_\infty-\xi j_\infty^2)^2\times
\begin{cases}
\Lambda^{-2+4\al}\;,&\al<1+\n/2\\
\Lambda^{2+2\n}\;,&\al>1+\n/2\;.
\end{cases}
\ee
From this expression one infers the suppression scale $M_4$. Restoring
the explicit dependence on $\Lambda_*$ one obtains,
\be
\label{M4Lif}
M_4\sim\begin{cases}
\Lambda_*\,\big(\Lambda_*/\Lambda\big)^{-1+2\al}\;,&\al<1+\n/2\\
\Lambda_*\,\big(\Lambda_*/\Lambda\big)^{1+\n}\;,&\al>1+\n/2\;.
\end{cases}
\ee
Note that $M_4$ is parametrically higher than $\Lambda_*$ for $\al>1/2$.

Case {\bf (ii)}: $\al>1+\n$. The matrix element is
dominated by the region $u\lesssim 1$. Thus, we use Eqs.~(\ref{phina}),
(\ref{anbn}), (\ref{bn1}) and obtain,
\[
\bra{\phi_0} V\ket{\phi_n}=k^2\cdot
\frac{2^{1-\n}\sqrt{1+\n}}{\Gamma(1+\n)}\,
\frac{\Lambda^{2+\n}m_n^{\n}}{|J_\n(m_n/\Lambda)|}\;B\;,
\] 
where $B$ is defined in (\ref{calI}). Substitution of this expression
into Eq.~(\ref{PTformulas}) produces a divergent sum, which we cut off
at $n_{max}$. The final result reads,
\be
\label{domega22}
\delta\omega_{0,(2)}^2\propto-k^4B^2\Lambda^{2+2\n}\;.
\ee
Note that the parametric dependence on the confinement scale $\Lambda$
is the same as in the lower case of
Eq.~(\ref{domega21}). Correspondingly, the mass parameter $M_4$
is given by the
lower case of (\ref{M4Lif}).

Let us discuss the conditions for the validity of the
quantum-mechanical perturbation theory used above. A necessary
requirement is that the correction to the eigenvalue is smaller than
the spacing between the adjacent levels of the unperturbed
Hamiltonian. Besides, the second-order correction must be smaller than
the first one. This amounts to the conditions,
\be
\label{PTconds}
\delta\omega_{0,(1)}^2\ll\Lambda^2~,~~~~~
\delta\omega_{0,(2)}^2\ll\delta\omega_{0,(1)}^2\;.
\ee
Clearly, they restrict the values of the spatial momentum $k$ where the
perturbative formulas make sense.
By inspection of (\ref{domega1n4}), (\ref{domega1n8}) we find that the
first inequality in (\ref{PTconds}) implies
\[
k\ll\min\{\Lambda^{1-\al},\Lambda^{-\n}\}\;,
\]
while Eqs.~(\ref{domega21}), (\ref{domega22}) and the second
inequality lead to 
\[
k\ll\begin{cases}
\Lambda^{1-\al}\;,&\al<1+\n/2\\
\Lambda^{-1-\n+\al}\;,&1+\n/2<\al<1+\n\\
1&\al>1+\n\;.
\end{cases}
\]
One observes that the bound following from the second order of the
perturbation theory is stronger for $\al>1+\n/2$. We conjecture that
considering even higher orders will lower the bound further in the
range $1<\al<1+\n$, so that eventually one will arrive at
\be
\label{kmax}
k\ll\min\{\Lambda^{1-\al},1\}.
\ee
In other words, the upper bound on the spatial momentum for which 
the perturbative
expansion of the dispersion relations works, is smaller or equal to the LV scale
$\Lambda_*$. This is quite similar to the situation in the RS-type
model of Sec.~\ref{sec:RS}.

\subsubsection{Zero-mode dispersion at large momenta}

One would like to go beyond the limitation (\ref{kmax}) and study the
dispersion relations at large momenta. In particular, it is
interesting to understand if deviations from LI remain small when the
momentum exceeds the LV scale $\Lambda_*$. It turns out that, as far
as the zero mode is concerned, this question can be easily answered in
the affirmative if the parameter $\xi$ is zero or
negative. Note that according to Eqs.~(\ref{deltaomega1}) the first
corrections to the velocities of the bound sates in this case are
positive meaning that these modes are `superluminal'. 

The argument is based on the variational theorem that states that the
average of a Hamiltonian over any function is larger than the energy
of the ground state. Consider first the total Hamiltonian $\hat
H_{(0)}+V$. Its lowest eigenvalue gives the frequency $\omega_0$ of
the zero mode. Thus we have,
\[
\omega_0^2-k^2\leq\bra{\phi}\hat H_{(0)}+V\ket{\phi}
\] 
for any function $\phi(u)$. 
Let us substitute here the ground state eigenfunction $\phi_0$ of the
unperturbed Hamiltonian $\hat H_{(0)}$. Recalling that the lowest
eigenvalue of the latter is zero we obtain,
\be
\label{upper}
\omega_0^2-k^2\leq\bra{\phi_0}V\ket{\phi_0}=\delta\omega_{0,(1)}^2\;,
\ee
where the r.h.s. has been computed in Eqs.~(\ref{domega1n4}),
(\ref{domega1n8}) for the cases $\al<1+\n$ and $\al > 1+\n$
respectively. We conclude that the deviation of the (phase) velocity
of the mode from unity is bounded from above by a small quantity
behaving as a positive power of the IR scale $\Lambda$.

To obtain a lower bound we use the property of the domain wall
solution mentioned in Sec.~\ref{sec:geometry} that $f(u)$, $j(u)$ are
monotonically decreasing functions. For $\xi\leq 0$ this implies that
the potential $V(u)$ is also monotonically decreasing. Thus, it always
exceeds  
\[
\label{Vmin}
V_{min}=k^2(2f_\infty-\xi j_\infty^2)\Lambda^{2\al}\;,
\] 
which is attained at $u=\Lambda^{-1}$. We now have the following chain
of relations,
\be
\label{lower}
\omega^2_0-k^2=\bra{\phi_{0}^{ex}}\hat
H_{(0)}+V\ket{\phi_{0}^{ex}}
\geq\bra{\phi_{0}^{ex}}\hat
H_{(0)}\ket{\phi_{0}^{ex}}+V_{min}\geq V_{min}\;,
\ee
where $\phi_{0}^{ex}$ is the exact ground state wavefunction of the
full Hamiltonian $\hat H_{(0)}+V$. In the last inequality we have
again applied the variational theorem, this time to the unperturbed
Hamiltonian $\hat H_{(0)}$. Eq.~(\ref{lower}) gives the desired lower
bound on the deviation of the mode velocity from one. We see that in
the case under study ($\xi\leq 0$) it is positive and proportional to
$\Lambda^{2\al}$. Combining the bounds (\ref{upper}) and (\ref{lower})
one concludes that for a large hierarchy between the IR scale
$\Lambda$ and the LV scale $\Lambda_*$ the dispersion relation of the
lightest bound state is confined within a narrow wedge close to the
lightcone.  

Interestingly, we see that the Lifshitz model 
has the same property as the Randall--Sundrum model of
Sec.~\ref{sec:RS}:  
the dispersion relation of the zero mode displays 
the emergent relativistic form at all momenta, including
$k\gg \Lambda_*$. Again, we observe that this property correlates with
the superluminality of the bounds states. Similarly to the
Randall--Sundrum case,  
this behaviour can be understood as due to 
the additional contribution to the potential \eqref{H1} 
representing
an effective $k-$dependent mass term peaked in the Lifshitz part of
the geometry. In the superluminal case the squared mass is positive 
so that the zero mode is more and
more repelled from the UV at large momenta, leading to 
the recovery of the emergent speed. The same reasoning suggests that
the persistence of LI at all momenta must also hold for the excited KK
modes, though we did not find a simple way to demonstrate this explicitly. 
To sum up, it seems rather generic that the emergence of LI is more
efficient for  
superluminal models, at least within the validity of
the holographic 
correspondence.

\section{Summary and Discussion}
\label{sec:discussion}

In this paper we have studied the emergence of Lorentz invariance (LI)
at low energies in strongly coupled Lorentz violating (LV)
scale-invariant field theories (SFT's). We have considered two
scenarios: relativistic CFT's which are perturbed at high energies by
coupling to an `external' LV sector; and Lifshitz SFT's which flow in
the IR to a LI fixed point. Using explicit models that provide the
holographic description of these scenarios, we have analyzed the
implications of the emergent LI for observable quantities, such as
correlators and dispersion relations of the bound states, and
developed a systematic approach to estimate the LV corrections. 
We have confirmed that the strong dynamics accelerates the
renormalization group (RG)  flow 
towards LI at low energies.
In
agreement with the general RG expectations, 
we found that the leading LV corrections are
power-law suppressed by the ratio of the IR scale to the scale of LV,
with the exponent related to the dimension of the {\it least
  irrelevant Lorentz violating operator} (LILVO). We have identified
the IR scale that must be substituted in the RG formulas. For the
correlation function in the Euclidean domain this 
is nothing but the inverse distance/time, at which the correlation is
measured. For the bound 
state spectrum the relevant IR scale coincides with 
the scale of confinement; this can be much lower than
the absolute energy or momentum of the particle. In other words, the
leading corrections to the dispersion relations have the form of the
shift in the propagation velocities, 
$
\omega^2\simeq (1+2\delta c)\,k^2$,
with 
\[
\delta c\propto \left({\Lambda / \Lambda_*}\right)^{\Delta-d}\;,
\]
where $\Lambda_*$ is the scale of high-energy LV, $\Lambda$ is the
confinement scale and $\Delta$ is the dimension of the LILVO. These
findings are common to both scenarios that we considered and we
believe them to be valid for any system where LI emerges due to strong
dynamics. 

On the other hand, the nature and dimension of LILVO is
model-dependent. In the Randall--Sundrum-type model realizing the
first scenario the role of LILVO is played by the kinetic term
$\dot{\bar\phi}^2$ of the `external' LV scalar field which couples to
a CFT through a scalar operator $O_\phi$. Close to the IR fixed point
$\bar\phi$ acquires an anomalous dimension, so that 
$\dim \bar\phi=d-\dim O_\phi$. Then for the dimension of LILVO we obtain 
\be
\label{Delta1}
\Delta_1=2+2d-2\dim O_\phi\;.
\ee 
The case $\dim O_\phi=d/2$, which is the
minimal dimension of the scalar operator achievable in this setup,
gives the most efficient recovery of LI with the leading corrections,
such as $\delta c$, suppressed by two powers
of $\Lambda/\Lambda_*$. 

It might seem
  that a model with dimension of $O_\phi$ lower than $d/2$ (but still
  larger than the unitarity bound $d/2-1$) would lead to a larger
  $\Delta$ and hence stronger suppression of LV corrections. However,
  this is not true. Such a model will always contain a LV operator
  constructed of $O_\phi$ itself, namely ${\dot{O}_\phi}^2$, whose
  dimension is 
\be
\label{Delta2}
\Delta_2=2+2\dim O_\phi\;.
\ee
This is smaller than
  (\ref{Delta1}) for $\dim O_\phi<d/2$. Thus ${\dot{O}_\phi}^2$
  becomes LILVO and the suppression is actually weakened. This is
  precisely what happens for the alternative interpretation of the
  model of Sec.~\ref{sec:RS} mentioned in footnote~\ref{foot:1}.   

More constraints arise for a CFT that emerges as the IR endpoint of a
Lifshitz flow. It contains the additional LV perturbation by the
irrelevant spin-one operator,
\be
\label{dla}
\delta {\cal L} = \frac{O_{\cal A}^t}{\Lambda_*^{\alpha}}\;,
\ee
with  $\alpha={\rm dim}(O^\mu_{\cal A}) -d>0$. 
The latter imprints a modified scaling in the 
observables. In particular, the LV correction in the speed of the
scalar bound states receives an additional contribution scaling
like\footnote{The fact that the exponent equals $2\alpha$ and not just
$\alpha$ is a consequence of the additional discrete symmetry $O_{\cal
A}^\mu\mapsto -O_{\cal A}^\mu$ appearing at the IR fixed point. Due to
this symmetry the perturbation (\ref{dla}) contributes 
to the dispersion relations
of the scalar bound states only
at the second order of conformal perturbation theory.} 
$\delta c\sim \left({\Lambda / \Lambda_*}\right)^{2\alpha}$, which
competes with the intrinsic correction that scales according to
(\ref{Delta2}). Whenever $\dim O_\phi$ is bigger than $d/2-1+\alpha$,
the leading correction comes with the exponent $2\alpha$ and becomes
independent of $\dim O_\phi$.   
In the Lifshitz flows of \cite{Kachru:2008yh,Braviner:2011kz} 
$\alpha$ cannot be too large\footnote{Otherwise, the same bulk spin 1
  field cannot give a relevant 
  deformation of the UV Lifshitz point.} ($\alpha\lesssim0.35$ in $d=4$)
 and therefore the LV effects are
not very strongly suppressed. In view of this, it would be interesting
to search for
other (holographic) models with UV Lifshitz scaling
where the dimension of LILVO would be higher.

We have also analyzed the
next-to-leading LV corrections to the dispersion relations 
--- the contributions into $\omega^2$
of the form $k^{2l}M_{2l}^{\,2-2l}$ with $l=2,3,\ldots$. We have found
that the mass scale $M_{2l}$ that suppresses the order-$l$ correction is
given by $\Lambda_*$ multiplied by a certain power of the ratio
$\Lambda/\Lambda_*$. This power is
related to the dimension of the LILVO in a non-trivial way; furthermore,
it is different for the terms of different order $l$. 
In general, it can be both positive or
negative depending on the model parameters, and correspondingly the
scale $M_{2l}$ can be lower of higher than $\Lambda_*$. It is unclear
at the moment if the scaling behavior of $M_{2l}$ can be determined
from the general RG arguments or if it is strongly model-dependent.

We have observed that the Taylor expansion of the dispersion relations
in the powers of momentum inevitably breaks down at momenta at most as
high as $\Lambda_*$. This happens even if the individual terms in the
expansion are suppressed. In the Randall--Sundrum type model of
Sec.~\ref{sec:RS} it were possible to analyze the dispersion
relations rather explicitly past this critical momentum and we found
that, quite surprisingly, they tend to relativistic form at high
 $k\gg \Lambda_*$ if the `external' LV sector is
superluminal. In the subluminal case the recovery of the relativistic
dispersion relations still holds for the excited bound states, while
the lightest (massless) state gets non-relativistic at high
momenta. For the model of the Lifshitz flow studied in
Sec.~\ref{sec:flow} the dispersion relations cannot be determined
analytically. Still, for the parameters in a certain range
corresponding again to superluminal type of LV, we were able to obtain
constraints on the frequency of the massless state which show that its
dispersion relation lies in a narrow wedge close to the light-cone. In
other words, its phase velocity $\omega/k$ is constrained to be close
to 1 at arbitrary $k$. It would be interesting to understand whether this
property --- persistence of almost LI dispersion relations at all
momenta --- is specific to the class of holographic models studied in this
paper or is more generic.

Our results have implications in two areas. First, the holographic
models of the type studied in this paper may be relevant for the
description of emergent LI in condensed matter systems. The phenomenon
of emergent LI has 
been known for a while in the condensed matter literature, 
see {\em e.g.} \cite{Vafek:2002jf,Lee:2002qza,Herbut:2009qb}, where
it has been analyzed with a variety of methods including $1/N$-,
$\epsilon$-expansions and the exact RG. However, as we discussed, the
emergence of LI becomes truly efficient only when the theory gets strongly
coupled, which renders the traditional perturbative methods
unreliable. The holographic duality provides a complementary
approach to assess the emergence of LI in strongly coupled condensed
matter systems, which allows a systematic calculation of the LV
corrections and their scaling exponents. 
Admittedly, an important
challenge along these lines is to construct holographic models that
would 
faithfully represent the essential properties of real materials.

A related phenomenon, which also seems to be ubiquitous in condensed
matter context, is emergence of rotational invariance (or
isotropy). Despite the fact that any material lacks rotational
invariance at the 
 atomic/molecular level, 
 many of them display an isotropic response in the
 continuum limit. Again, the approach to the isotropic fixed point can
 be described using the holographic methods by an analytic continuation
 of our discussion\footnote{We actually performed such continuation
   when we studied the correlation functions.}. It would be
 interesting to establish the connection between the results of this
 paper and the previous studies of the emergence of isotropy in 
$O(N)$ models \cite{Campostrini} and
 lattice field theories \cite{Davoudi:2012ya}. In particular, it
 was found in these references that
in the continuum limit all anisotropic corrections scale 
with the value of the exponent close to~2. 
It would be interesting to see what is the relation  between this
and our findings regarding the optimal LILVO dimension in the case of the
Randall--Sundrum type model. 

Another important application of the mechanism
for emergent LI studied in this paper appear
in non-relativistic quantum gravity scenarios
\cite{Horava:2009uw,Blas:2009qj}.  
As mentioned in the Introduction, the idea (as in technicolor/compositeness models)
is to embed the Standard Model (SM) fields as the lightest
excitations of some strongly coupled sector which has order-one
LV in the UV. The strong
dynamics would then enforce the
approximate LI seen in the known particle physics.  
This requires realizing the SM fields as (partially- or completely-)
composite states. The most minimal holographic realization of a
completely composite SM in this context is the Randall-Sundrum I model
\cite{Randall:1999ee}, with the entire SM living on the IR brane and
LV physics restricted to the UV brane. In this model there is no tower
of states with the quantum numbers of the SM fields, so 
at leading order in $1/N$ there is no LV whatsoever in the
SM. Only the inescapable tower of spin 2 resonances receive
small LV corrections at low momenta in this approximation.  
Of course, quantum effects in the bulk are expected introduce some
$1/N$-suppressed LV into the SM sector. We leave 
the interesting question of estimating these effects for future.  

In order to be more generic, one can place the
SM fields in the bulk, realizing in this way `partial compositeness'. 
This will require to generalize the construction of the present paper
to include fermions and gauge fields. While inclusion of fermions is
expected to be straightforward and closely follow the track presented
in this paper for scalars\footnote{Fermions can be given
  different degree of compositeness by adjusting the bulk and brane
  mass terms~\cite{Contino:2004vy}. In the present context, since the
  deviations from LI are most tightly constrained for the first
  generation of quarks and leptons, these are
   the fermions that must be the most
  composite.}, one expects 
some qualitative differences to arise for the
gauge fields.
From the bulk perspective, the reason is that the
gauge-field zero-modes are delocalized in the bulk: their wave
functions penetrate into the UV and so they will be sensitive
to the LV UV physics. 
In the CFT language the gauge field in the bulk is dual to an
`elementary' 4-dimensional gauge field\footnote{To give rise to a
  zero-mode the bulk gauge field must obey Neumann boundary conditions
on the UV brane. Its boundary value is then interpreted as the
`elementary' field \cite{ArkaniHamed:2000ds}.} 
coupled to the CFT through a
conserved current with protected dimesion, $\dim J^\mu=3$. 
Such coupling is marginal and implies logarithmic sensitivity to the
UV. 
We have
observed this kind of behavior in the model of Sec.~\ref{sec:RS}
when the `elementary' scalar couples to the CFT via operator
of dimension $d-1$ (see the end of Secs.~\ref{ssec:propel},
\ref{ssec:RSbound}). In this case the LV corrections are indeed
suppressed only logarithmically.
This represents a 
serious obstruction, since it leads to the reappearance of the Lorentz
fine-tuning 
problem.

Let us point out a possible way to circumvent
this problem. It relies on the observation that even though
the LV corrections log-persist along the RG flow, they are inversely
proportional to the
coupling constant. Namely, the logarithms enter into the formulas in the 
combination
$g_4^{-2}(\log{\Lambda_*/\Lambda})^{-1}$ with $g_4\sim \sqrt{L/b}$ being
the coupling between the 4-dimensional gauge field and the CFT
(cf. Eqs.~(\ref{fsprop3}), (\ref{lightmode3})). 
This suggests that there
should still be a limit where the LV corrections are suppressed,
involving again strong coupling, $g_4\gg1$. 
In hindsight, one realizes that this 
limit corresponds to enhancing the gauge
kinetic term in the IR part of the geometry. A dynamical
realization of this might
be possible by introducing interaction of the bulk gauge field with 
a dilaton, see
\cite{Kehagias:2000au}. 
In this case the 4-dimensional gauge boson would become composite and the gauge
symmetry would be itself emergent. 
This scenario deserves further
study and is left for future investigation. 

Similarly, we can foresee what happens to gravity in this framework by
looking at the model of Sec.~\ref{sec:RS} in the case when the scalar
operator $O_\phi$ has dimension 4 (for $d=4$). Since this leads to an
irrelevant coupling, LI does not emerge in this 
sector. 
This property may be actually an advantage. It implies that it is
possible to have sizable LV in gravity without destroying the
stability of LI in the
matter sector. This is
interesting from the phenomenological viewpoint as it opens an 
opportunity to probe gravitational LV. The existing data already place
non-trivial bounds on LV in
gravity \cite{Will:2005va}. However, these bounds are much milder than in the
matter sector,  
and so only a
moderate amount of tuning is required to satisfy them.

There is one concern about embedding gravity in the picture.  The standard way to include gravity in this context is by introducing a UV brane that cuts off the UV end of the geometry, giving rise to a (quasi-)localized graviton mode. For Lifshitz geometries, this requires the UV brane to violate the Null Energy Condition (NEC) -- see  Appendix \ref{app:IRbrane} (cf. \cite{Dubovsky:2001fj,Gubser:2008gr}). However, this may not be a problem if the NEC-violating sector confined on the brane is not itself Lorentz invariant. Of course, this issue deserves further scrutiny.

Finally, let us emphasize that the existence of the strong-coupling
mechanism for emergence of LI 
together with the previously known ones (based on a non-relativistic
SUSY \cite{GrootNibbelink:2004za} and on a separation of scales
\cite{Pospelov:2010mp}) opens up an interesting connection between
particle physics phenomenology and non-relativistic
gravity. Generally, avoiding the fine-tuning problems associated
with fundamental LV will require some mechanism --- some
`new physics' --- operating down to a very low energy scale. Then via
naturalness the non-relativistic gravity models become very
predictive. A very rough estimate of the typical new-physics scale is
around 10 orders of magnitude below the LV scale. The latter is
bounded from above by the gravitational observations.
Taking it to be $\sim 10^{15}$ GeV \cite{Blas:2009ck,Blas:2010hb},
one obtains a new physics scale as low as $\sim 100$ TeV. By improving the
constraints on the new physics scale on one hand and on the LV scale
on the other, one may eventually cover the whole 
window required for the
mechanisms of emergent LI to operate. In our opinion, this further
motivates  the experimental searches for new physics,
including Lorentz Violation.

\paragraph*{Acknowledgements}

We are indebted to Riccardo Rattazzi for sharing his insight in the
properties of Lorentz invariant fixed points. We are grateful to
Dmitry Levkov, Alex Pomarol, Marco Serone and Andrea Wulzer for useful
discussions. We also thank the organizers and participants of the
Kavli IPMU Focus Week on Gravity and Lorentz Violations for the
encouraging interest and valuable comments. 
S.S. thanks the Center for
Cosmology and Particle Physics of NYU for hospitality during the completion
of this work.  
S.S. is supported by 
the Grant of the President of Russian Federation
NS-5590.2012.2, the Russian Ministry of Science and Education
under the contract 8412, the RFBR grants 
11-02-01528, 12-02-01203
and by the Dynasty Foundation. 
O.P. is supported by a Ram\'on y Cajal fellowship (subprograma MICINN-RYC),
and MICINN under contract FPA 2011-25948.

\appendix

\section{Bound states in RS model with irrelevant coupling}
\label{app:irr}

Here we consider the model of Sec.~\ref{sec:RS} with the
choice of parameters such that $\n>1$. According to the discussion in
the main text, in this case one does not expect appearance of LI at
low energies. We confirm this expectation explicitly by studying
the spectrum of KK modes.

The KK spectrum is still determined by the basic
Eq.~(\ref{AdSdet}). Let us restrict to the choice
$\m_{UV}=\m_{IR}=0$ and make a technical assumption that $\n$
is non-integer. This will simplify manipulations with the Bessel
functions. Expanding the contributions with the ``UV''-subscript at
$pL\ll 1$ one obtains from (\ref{AdSdet}),
\be
\label{Jexp3}
\frac{(1+\vk)\omega^2-(c^2+\vk)k^2}{\vk\L^2}
\bigg(\frac{p}{2\L}\bigg)^{-2\n} J_{\n-1}(p/\L)=
\frac{4\Gamma(2-\n)}{\Gamma(\n)}(L\L)^{2(\n-1)}
J_{1-\n}(p/\L)\;,
\ee
where $\vk$ has been defined in (\ref{vk}).
The r.h.s. here is suppressed by the small factor
$(L\L)^{2(\n-1)}$. Neglecting this contribution one obtains a LV
gapless mode with the dispersion relation 
\be
\label{LVmode}
\omega^2=\frac{c^2+\vk}{1+\vk}\,k^2\;
\ee  
and a family of LI massive modes,
\be
\label{pfamily}
\omega_n^2=k^2+(\L j_{\n-1}^{(n)})^2\;,~~~n=1,2,\ldots\;,
\ee
where $j_{\n-1}^{(n)}$ are positive roots of $J_{\n-1}$. The formula
(\ref{LVmode}) clearly matches with the two-point function of
$\bar\phi$, Eq.~(\ref{fsprop1}), and thus the corresponding mode is
identified with the elementary LV scalar whose velocity has been
renormalized by the interaction with the CFT. On the other hand, 
the family (\ref{pfamily}) gives just the standard KK spectrum 
of the bulk scalar in AdS with the Dirichlet boundary conditions on
the UV brane. So from the dual viewpoint these are the bound states of
the CFT decoupled from the LV deformation. 

It remains to see how this picture is affected by taking into account
the r.h.s. of (\ref{Jexp3}). One finds that the frequencies of the CFT
modes get shifted by 
\be
\label{deltaomega}
\delta\omega^2_n=\L^2\cdot\frac{2\Gamma(2-\n)}{\Gamma(\n)}
\cdot\frac{\big(j_{\n-1}^{(n)}\big)^{2\n+1} J_{1-\n}\big(j_{\n-1}^{(n)}\big)}
{J'_{\n-1}\big(j_{\n-1}^{(n)})}
\cdot\frac{\vk (L\L)^{2(\n-1)}}
{(1-c^2)k^2+(1+\vk)\big(\L j_{\n-1}^{(n)}\big)^2}\;.
\ee
For $c^2<1$ this is basically the end of the story: the corrections
(\ref{deltaomega}) are uniformly small at all spatial momenta, so the
spectrum indeed splits into almost LI CFT plus a single LV mode.

However, if $c^2>1$ the corrections blow up at
\be
\label{kblowup}
k^2=\frac{(1+\vk)}{c^2-1}\big(\L j_{\n-1}^{(n)}\big)^2
\ee
implying that close to these points the perturbative scheme breaks
down. To understand what happens one observes that (\ref{kblowup})
corresponds to the intersections between the dispersion relations
(\ref{LVmode}) and (\ref{pfamily}). 
\begin{figure}[tb]
\begin{center}
\begin{picture}(400,120)(20,25)
\put(10,25){\includegraphics[scale=0.7]{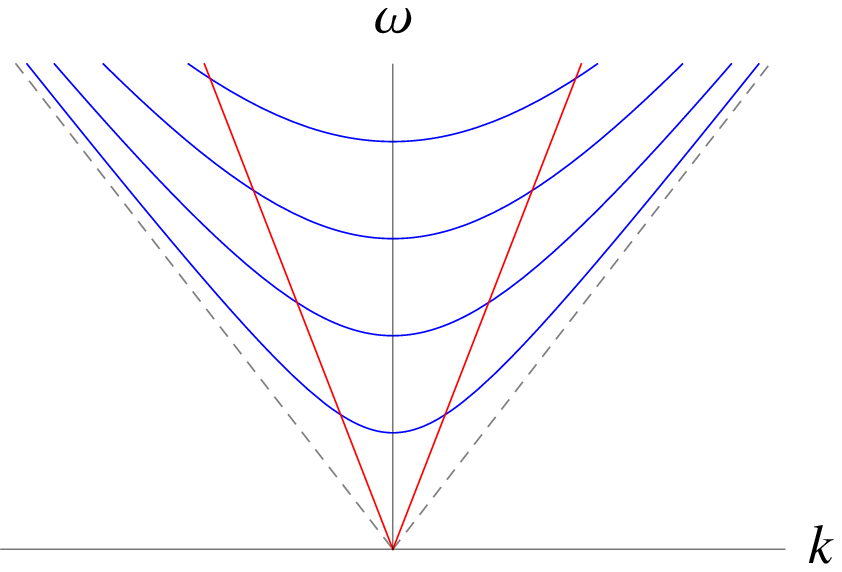}}
\put(210,80){$\Longrightarrow$}
\put(270,25){\includegraphics[scale=0.7]{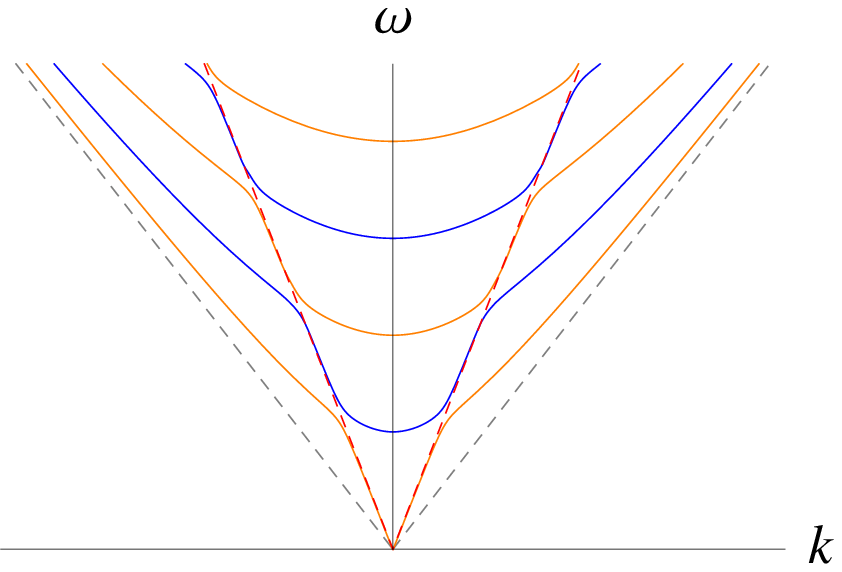}}
\put(85,5){(a)}
\put(345,5){(b)}
\end{picture}
\end{center}
\caption{Dispersion relations of the eigenmodes in the leading
  approximation (a) and after taking into account the mixing (b).}
\label{Fig:1}
\end{figure}
Thus we have a situation analogous
to the level-crossing in quantum mechanics:
as far as we completely neglect
the r.h.s. in (\ref{Jexp3}) there are two
eigenvalues (frequencies) that cross at a given value of spatial
momentum. It is known that the perturbation actually prevents the levels
from crossing, so that the dispersion relations of the modes are
reconnected, see Fig.~\ref{Fig:1}. This produces a rather
peculiar situation when the LV mode (\ref{LVmode}) is ``cut in
pieces'' and its chunks are inserted between the LI KK modes. As a
result we obtain a tower of modes, all of which are essentially LI at
large momentum, while the massive modes are also LI at small $k$.
However, every mode strongly violates Lorentz invariance in some range
of momenta (for the gapless mode this range includes $k=0$), so this
case clearly does not exhibit emergent LI.

\section{Lifshitz flow: Corrections to the correlator in IR}
\label{app:corr}

Un this Appendix we emplement the perturbative scheme described in
Sec.~\ref{sec:correlators} to determine the LV corrections to the
correlator (\ref{corr4}). The reader should refer to that section for
the notations and conventions.

\subsection{Solutions in the $a'$-region}

We start with the $a'$-region where we need to find the solutions to
Eq.~(\ref{scalareq}) up to the order $O(w^2,k^2)$. Treating the terms
containing $w^2,k^2$ in (\ref{scalareq}) as a perturbation, one uses
the standard ``variation of constants'' technique and writes,
\be
\label{phiccorr}
\phi_{c,\sigma}(u)=P_{\sigma\tau}(u)\phi_{c,\tau}^{(0)}(u)\;,
\ee
where the indices $\sigma,\tau$ take the values $\pm$ and
the coefficient functions are required to satisfy,
\[
P_{\sigma\tau}'\phi_{c,\tau}^{(0)}=0\;,~~~~~~
P_{\sigma\tau}'{\phi_{c,\tau}^{(0)}}'=
g^2\bigg(\frac{1+\xi j^2}{f^2}w^2+k^2\bigg)\phi_{c,\sigma}^{(0)}\;.
\]
This yields,
\bseq
\label{PRs}
\begin{align}
P_{--}(u)&=1-\int_{u_0}^u\frac{d\tilde u}{2\nu}{\cal K}(\tilde u)
\phi_{c,-}^{(0)}\phi_{c,+}^{(0)}\;,
&P_{-+}(u)=\int_{u_0}^u\frac{d\tilde u}{2\nu}{\cal K}(\tilde u)
\big(\phi_{c,-}^{(0)}\big)^2\;,\\
P_{+-}(u)&=-\int_{u_0}^u\frac{d\tilde u}{2\nu}{\cal K}(\tilde u)
\big(\phi_{c,+}^{(0)}\big)^2\;,
&P_{++}(u)=1+\int_{u_0}^u\frac{d\tilde u}{2\nu}{\cal K}(\tilde u)
\phi_{c,+}^{(0)}\phi_{c,-}^{(0)}\;,
\end{align}
\eseq
where
\[
{\cal K}(u)\equiv u^{1-d}g\bigg(\frac{1+\xi
  j^2}{f}w^2+fk^2\bigg)\;,
\]
and $u_0$ is an arbitrary normalization point in the $c$-region,
i.e. $1\ll u_0\ll w^{-1},k^{-1}$. In deriving (\ref{PRs}) we have used
the expression for the Jacobian of the functions $\phi^{(0)}_{c,-}$,  
$\phi^{(0)}_{c,+}$,
\[
W[\phi_{c,+}^{(0)},\phi_{c,-}^{(0)}]=2\nu \frac{g u^{d-1}}{f}\;, 
\]
that follows from their asymptotics (\ref{phic0}).

Similarly, for the solutions defined by the asymptotics (\ref{phias})
we obtain,
\bseq
\label{phiacorr}
\begin{align}
\label{phia-corr}
\phi_{a,-}(u)&=\phi_{a,-}^{(0)}(u)\bigg(1-
\int_{0}^u\frac{g_-d\tilde u}{f_0(\lambda_+-\lambda_-)}
{\cal K}\phi_{a,-}^{(0)}\phi_{a,+}^{(0)}\bigg)
+\phi_{a,+}^{(0)}(u)\int_{u_1}^u\frac{g_-d\tilde u}{f_0(\lambda_+-\lambda_-)}
{\cal K}\,
\big(\phi_{a,-}^{(0)}\big)^2\;,\\
\label{phia+corr}
\phi_{a,+}(u)&=
-\phi_{a,-}^{(0)}(u)
\int_{0}^u\frac{g_-d\tilde u}{f_0(\lambda_+-\lambda_-)}
{\cal K}\,\big(\phi_{a,+}^{(0)}\big)^2
+\phi_{a,+}^{(0)}(u)\bigg(1+
\int_{0}^u\frac{g_-d\tilde u}{f_0(\lambda_+-\lambda_-)}
{\cal K}\phi_{a,+}^{(0)}\phi_{a,-}^{(0)}\bigg)\;,
\end{align}
\eseq
where we used the Jacobian
\[
W[\phi_{a,+}^{(0)},\phi_{a,-}^{(0)}]
=\frac{f_0(\lambda_+-\lambda_-)}{g_-}\frac{gu^{d-1}}{f}\;.
\]
Note that all but one integrals appearing above converge at $\tilde
u\to 0$ which allowed to extend the domain of integration down to this
point. On the other hand, the integral entering the second term in
(\ref{phia-corr}) diverges at $\tilde u\to 0$, so we have normalized it at
an arbitrarily chosen point $u_1\ll 1$. We will see below that the
final answer does not depend on this choice.
Combining (\ref{phiacorr}) with (\ref{phiccorr}), (\ref{PRs}) one
arrives at the analog of the relations (\ref{phica0}), 
\[
\phi_{c,\sigma}=(V_{\sigma\tau}^{(0)}+{\cal I}_{\sigma\tau})\phi_{a,\tau}
\]
with
\begin{align}
&{\cal I}_{--}=\int_{0}^{u_0}\frac{g_-d\tilde
  u}{f_0(\lambda_+-\lambda_-)}
{\cal K}\phi_{c,-}^{(0)}\phi_{a,+}^{(0)}\;,\notag\\
&{\cal I}_{-+}=-\int_{u_1}^{u_0}\frac{g_-d\tilde
  u}{f_0(\lambda_+-\lambda_-)}
{\cal K}\phi_{c,-}^{(0)}\phi_{a,-}^{(0)}
-V^{(0)}_{-+}\int_{0}^{u_1}\frac{g_-d\tilde
  u}{f_0(\lambda_+-\lambda_-)}
{\cal K}\phi_{a,-}^{(0)}\phi_{a,+}^{(0)}\;,\notag\\
&{\cal I}_{+-}=\int_{0}^{u_0}\frac{g_-d\tilde
  u}{f_0(\lambda_+-\lambda_-)}
{\cal K}\phi_{c,+}^{(0)}\phi_{a,+}^{(0)}\;,\notag\\
&{\cal I}_{++}=-\int_{u_1}^{u_0}\frac{g_-d\tilde
  u}{f_0(\lambda_+-\lambda_-)}
{\cal K}\phi_{c,+}^{(0)}\phi_{a,-}^{(0)}
-V^{(0)}_{++}\int_{0}^{u_1}\frac{g_-d\tilde
  u}{f_0(\lambda_+-\lambda_-)}
{\cal K}\phi_{a,-}^{(0)}\phi_{a,+}^{(0)}\;,\notag
\end{align}
In deriving these expressions we made use of the formula for the
determinant of the matrix~$V_{\pm\pm}^{(0)}$,
\[
\big|V^{(0)}\big|=\frac{W[\phi_{c,+}^{(0)},\phi_{c,-}^{(0)}]}
{W[\phi_{a,+}^{(0)},\phi_{a,-}^{(0)}]}=\frac{2\nu
  g_-}{f_0(\lambda_+-\lambda_-)}\;. 
\]

\subsection{Solution in the $b$-region}

In the $b$-region one has to take into account the deviation of the
background functions from the constant values, see
Eqs.~(\ref{coeffs}). The solution is again found by the ``variation of
constants'': 
\[
\phi_{b}(u)=R(u)\phi_{b}^{(0)}(u)+Q(u)\phi_{b,+}^{(0)}(u)\;,
\]
where $\phi_{b}^{(0)}$ is given by (\ref{phib-0}) and 
\[
\phi_{b,+}^{(0)}=u^{d/2}I_\nu(p_E u)
\]
is the second linearly independent solution of the unperturbed equation. The
coefficient functions must satisfy
\[
R'\phi_{b}^{(0)}+Q'\phi_{b,+}^{(0)}=0\;,~~~~
R'{\phi_{b}^{(0)}}'+Q'{\phi_{b,+}^{(0)}}'={\cal F}(u)\;,
\]
where
\[
{\cal F}(u)=
2\alpha(f_\infty-g_\infty)u^{-2\alpha-1}{\phi_{b}^{(0)}}'\notag\\
+\big[\big((\xi
j_\infty^2-2f_\infty+2g_\infty)w^2+2g_\infty k^2\big)u^{-2\alpha}
+2g_\infty(\mu L)^2u^{-2\alpha-2}\big]\phi_{b}^{(0)}\;.
\]
This yields
\be
\label{RQ}
R(u)=1-\int_{u_0}^ud\tilde u\,
{\tilde u}^{1-d}{\cal F}(\tilde u)\phi_{b,+}^{(0)}(\tilde u)\;,~~~~~
Q(u)=\int_{\infty}^ud\tilde u\,
{\tilde u}^{1-d}{\cal F}(\tilde u)\phi_{b}^{(0)}(\tilde u)
\ee
where we have used the Jacobian 
\[
W[\phi_{b,+}^{(0)},\phi_{b}^{(0)}]=u^{d-1}~,~~~
\]
Note that the
integral in the expression for $Q(u)$ is taken from infinity to ensure
that the solution vanishes at $u\to \infty$.

\subsection{Matching solutions in the $c$-region}

We need to relate the solution $\phi_b$ to $\phi_{c,\pm}$,
\[
\phi_{b}(u)=U_-\phi_{c,-}(u)+U_{+}\phi_{c,+}(u)\;,
\]
where $U_\pm$ are constants. To this end we expand $\phi_b$,
$\phi_{c,\pm}$ in the $c$-region in the powers of $u$ and match the
corresponding coefficients. It will be enough to keep track only of
the terms containing $u^{d/2\pm\nu}$. 

We start with the expansion of $\phi_{c,\pm}$. 
To evaluate the
integrals entering (\ref{PRs}) one uses the asymptotic expressions
at $u\gg 1$,
\begin{align}
&{\cal K}(u)\approx u^{1-d}\big[
\big(1+(g_\infty+f_\infty)u^{-2\al}\big)p_E^2
+(\xi j_\infty^2-2f_\infty)u^{-2\al} w^2\big]\;,\notag\\
&\phi_{c,\pm}^{(0)}\approx u^{d/2\pm\nu}
(1+a_{\pm}u^{-2\alpha})~,~~~~
a_\pm=\frac{\alpha(d\pm 2\nu)(f_\infty-g_\infty)
+2g_\infty(\mu L)^2}{4\al(\al\mp\nu)}\;.\notag
\end{align}
Note that one should be careful to take into
account the subleading terms arising from the tails of the domain
wall background. Substitution into (\ref{PRs}), (\ref{phiccorr})
yields,
\bseq
\label{phicexpand}
\begin{align}
\phi_{c,-}&=u^{d/2-\nu}\bigg\{1+\frac{u_0^2p_E^2}{4\nu}
+\frac{u_0^{2-2\al}}{4\nu(1-\al)}\big[(g_\infty+f_\infty+a_++a_-)p_E^2
+(\xi j_\infty^2-2f_\infty)w^2\big]\bigg\}\notag\\
&-u^{d/2+\nu}\bigg\{\frac{u_0^{2-2\nu}p_E^2}{4\nu(1-\nu)}
+\frac{u_0^{2-2\nu-2\al}}{4\nu(1-\nu-\al)}
\big[(g_\infty+f_\infty+2a_-)p_E^2+(\xi
j_\infty^2-2f_\infty)w^2\big]\bigg\}
+\ldots\;,\\
\phi_{c,+}&=u^{d/2-\nu}\bigg\{\frac{u_0^{2+2\nu}p_E^2}{4\nu(1+\nu)}
+\frac{u_0^{2+2\nu-2\al}}{4\nu(1+\nu-\al)}
\big[(g_\infty+f_\infty+2a_+)p_E^2+(\xi j_\infty^2-2f_\infty)w^2\big]
\bigg\}\notag\\
&+u^{d/2+\nu}\bigg\{1-\frac{u_0^2p_E^2}{4\nu}
-\frac{u_0^{2-2\al}}{4\nu(1-\al)}\big[(g_\infty+f_\infty+a_++a_-)p_E^2
+(\xi j_\infty^2-2f_\infty)w^2\big]
\bigg\}+\ldots\;,
\end{align}
\eseq
where dots stand for the terms with other powers of $u$.

To find the relevant terms in the expansion of the function $\phi_b$
one makes two observations. First, the constant shift of
the function $R(u)$, see (\ref{RQ}), amounts to the change of the
overall normalization of $\phi_b$ and thus will not affect the result
for the correlator (which is of course independent of this
normalization). It is straightforward to see that only the constant part of
$R(u)$ contributes into the coefficients of interest. We will
conveniently normalize this contribution to 1. Second, the function
$Q(u)$ can be evaluated explicitly using the formula,
\[
\begin{split}
\int_y^\infty dx\,&x^{-1-2\alpha}\big(K_\nu(x)\big)^2=
\frac{\sqrt\pi\Gamma(-\al)\Gamma(-\al-\n)\Gamma(-\al+\n)}
{4\Gamma(1/2-\al)}\\
&-\frac{\pi y^{-2\al}}{4\n\al\sin\pi\n}
\;{}_2F_3\bigg(\frac{1}{2},-\al;1-\al,1-\n,1+\n;y^2\bigg)\\
&+\frac{2^{2\n} y^{-2\al-2\n}}{8(\al+\n)}(\Gamma(\n))^2
\;{}_2F_3\bigg(\frac{1}{2}-\n,-\al-\n;1-\al-\n,1-2\n,1-\n;y^2\bigg)\\
&+\frac{2^{-2\n} y^{-2\al+2\n}}{8(\al-\n)}(\Gamma(-\n))^2
\;{}_2F_3\bigg(\frac{1}{2}+\n,-\al+\n;1-\al+\n,1+2\n,1+\n;y^2\bigg)\;,
\end{split}
\]
where ${}_2F_3$ is the generalized hypergeometric function. 
It is an entire function of its last argument
and thus adds only integer powers of $y^2$. Given this information, we
obtain,
\be
\label{phibexpand}
\begin{split}
\phi_b=U_-^{(0)} u^{d/2-\n}
+U_+^{(0)} u^{d/2+\n}\Big[&1+s_\al\big(2g_\infty p_E^2+
(\xi j_\infty^2-2f_\infty)w^2\big)p_E^{-2+2\al}\\
&+s_{1+\al}\big(\al(d+2\al)(f_\infty-g_\infty)+2g_\infty(\m L)^2\big)
p_E^{2\al}\Big]+\ldots\;,
\end{split}
\ee
where $U_\pm^{(0)}$ are given in (\ref{U0s}) and
\[
s_\al=\frac{\sin\pi\n\,\Gamma(1-\al)\Gamma(1-\al-\n)\Gamma(1-\al+\n)}
{2\sqrt{\pi}\Gamma(3/2-\al)}\;.
\]
Combining (\ref{phibexpand}), (\ref{phicexpand}) we find,
\begin{align}
U_-=&U_-^{(0)}\bigg\{1-\frac{u_0^2p_E^2}{4\nu}
-\frac{u_0^{2-2\al}}{4\nu(1-\al)}\big[(g_\infty+f_\infty+a_++a_-)p_E^2
+(\xi j_\infty^2-2f_\infty)w^2\big]\bigg\}\notag\\
-&U_+^{(0)}\bigg\{\frac{u_0^{2+2\nu}p_E^2}{4\nu(1+\nu)}
+\frac{u_0^{2+2\nu-2\al}}{4\nu(1+\nu-\al)}
\big[(g_\infty+f_\infty+2a_+)p_E^2+(\xi j_\infty^2-2f_\infty)w^2\big]
\bigg\}\notag\\
U_+=&U_-^{(0)}\bigg\{\frac{u_0^{2-2\nu}p_E^2}{4\nu(1-\nu)}
+\frac{u_0^{2-2\nu-2\al}}{4\nu(1-\nu-\al)}
\big[(g_\infty+f_\infty+2a_-)p_E^2+(\xi
j_\infty^2-2f_\infty)w^2\big]\bigg\}\notag\\
+&U_+^{(0)}\bigg\{1+\frac{u_0^2p_E^2}{4\nu}
+\frac{u_0^{2-2\al}}{4\nu(1-\al)}\big[(g_\infty+f_\infty+a_++a_-)p_E^2
+(\xi j_\infty^2-2f_\infty)w^2\big]\notag\\
&+s_\al\big(2g_\infty p_E^2+
(\xi j_\infty^2-2f_\infty)w^2\big)p_E^{-2+2\al}
+s_{1+\al}\big(\al(d+2\al)(f_\infty-g_\infty)+2g_\infty(\m L)^2\big)
p_E^{2\al}\bigg\}\notag\;.
\end{align}

\subsection{Evaluation of the correlator}

We are now ready to evaluate the coefficients $T_\pm$ entering
(\ref{phiba}). A simple algebra yields,
\[
T_\sigma=U_\tau^{(0)}(V_{\tau\sigma}^{(0)}+\bar{\cal I}_{\tau\sigma})\;,
\]
where keeping only the LV contributions,
\bseq
\label{barIs}
\begin{align}
\bar{\cal I}_{--}=w^2\bigg\{\int_0^{u_0}\!\!du\,
\bar{\cal K}\phi_{c,-}^{(0)}\phi_{a,+}^{(0)}
-(\xi
j_\infty^2-2f_\infty)\bigg[V_{--}^{(0)}\frac{u_0^{2-2\al}}{4\n(1-\al)}
-V_{+-}^{(0)}&\frac{u_0^{2-2\n-2\al}}{4\n(1-\n-\al)}\bigg]\bigg\}
+\ldots\;,\\
\bar{\cal I}_{+-}=w^2\bigg\{\int_0^{u_0}\!\!du\,
\bar{\cal K}\phi_{c,+}^{(0)}\phi_{a,+}^{(0)}
+(\xi
j_\infty^2-2f_\infty)\bigg[V_{+-}^{(0)}\frac{u_0^{2-2\al}}{4\n(1-\al)}
-V_{--}^{(0)}&\frac{u_0^{2+2\n-2\al}}{4\n(1+\n-\al)}\notag\\
+&V_{+-}^{(0)}s_\al p_E^{-2+2\al}\bigg]\bigg\}
+\ldots\;,
\end{align}
\begin{align}
\bar{\cal I}_{-+}=w^2\bigg\{&-\int_{u_1}^{u_0}\!\!du\,
\bar{\cal K}\phi_{c,-}^{(0)}\phi_{a,-}^{(0)}
-V_{-+}^{(0)}\int_{0}^{u_1}\!\!du\,
\bar{\cal K}\phi_{a,-}^{(0)}\phi_{a,+}^{(0)}\notag\\
&-(\xi
j_\infty^2-2f_\infty)\bigg[V_{-+}^{(0)}\frac{u_0^{2-2\al}}{4\n(1-\al)}
-V_{++}^{(0)}\frac{u_0^{2-2\n-2\al}}{4\n(1-\n-\al)}\bigg]\bigg\}
+\ldots\;,\\
\bar{\cal I}_{++}=w^2\bigg\{&-\int_{u_1}^{u_0}\!\!du\,
\bar{\cal K}\phi_{c,+}^{(0)}\phi_{a,-}^{(0)}
-V_{++}^{(0)}\int_{0}^{u_1}\!\!du\,
\bar{\cal K}\phi_{a,-}^{(0)}\phi_{a,+}^{(0)}\notag\\
&+(\xi
j_\infty^2-2f_\infty)\bigg[V_{++}^{(0)}\frac{u_0^{2-2\al}}{4\n(1-\al)}
-V_{-+}^{(0)}\frac{u_0^{2+2\n-2\al}}{4\n(1+\n-\al)}
+V_{++}^{(0)}s_\al p_E^{-2+2\al}
\bigg]\bigg\}
+\ldots\;,
\end{align}
\eseq
and 
\[
\bar{\cal K}(u)=\frac{g_- u^{1-d}}{f_0(\lambda_+-\lambda_-)}\;
g\bigg(\frac{1+\xi j^2}{f}-f\bigg)\;.
\]
Note that expressing $\phi_{a,\sigma}^{(0)}$ via
$\phi_{c,\sigma}^{(0)}$ by inverting the relations (\ref{phica0}) one
can check that $\bar{\cal I}_{\sigma\tau}$ are independent of the
normalization point $u_0$ within the accuracy of our approximation, as
it should be. On the other hand, the dependence on $u_1$, that affects
the definition of the solution $\phi_{a,-}$, is still present; it will
drop out only at the last step of the calculation.   

Finally, we must compute the ratio $T_+/T_-$ that according to
Eq.~(\ref{corr11}) will give us the correlator of interest. Expanding
this ratio to quadratic order in $U_+^{(0)}/U_-^{(0)}$ and to linear
order in
$\bar{\cal I}_{\sigma\tau}$ and dropping the
local contributions polynomial in $w^2$ and $k^2$ we find,
\[
\begin{split}
{\cal G}_\phi\propto\frac{U_+^{(0)}}{U_-^{(0)}}
\frac{\big|V^{(0)}\big|}{\big(V_{--}^{(0)}\big)^2}
\bigg[1+\frac{1}{\big|V^{(0)}\big|}\bigg(
V_{--}^{(0)}\bar{\cal I}_{++}-V_{-+}^{(0)}\bar{\cal I}_{+-}
-V_{+-}^{(0)}\bar{\cal I}_{-+}-V_{++}^{(0)}\bar{\cal I}_{--}
+\frac{2V_{+-}^{(0)}V_{-+}^{(0)}}{V_{--}^{(0)}}\bar{\cal
  I}_{--}\bigg)\\
+\frac{U_+^{(0)}}{U_-^{(0)}}\frac{1}{\big|V^{(0)}\big|}\bigg(
-V_{+-}^{(0)}\bar{\cal I}_{++}-V_{++}^{(0)}\bar{\cal I}_{+-}
+\frac{2V_{++}^{(0)}V_{+-}^{(0)}}{V_{--}^{(0)}}\bar{\cal
  I}_{--}
+\frac{2V_{+-}^{(0)}V_{-+}^{(0)}}{V_{--}^{(0)}}\bar{\cal
  I}_{+-}\\
+\frac{\big(V_{+-}^{(0)}\big)^2}{V_{--}^{(0)}}\bar{\cal
  I}_{-+}
-\frac{3\big(V_{+-}^{(0)}\big)^2V_{-+}^{(0)}}{\big(V_{--}^{(0)}\big)^2}
\bar{\cal
  I}_{--}\bigg)\bigg]\;.
\end{split}
\]
Inserting here Eqs.~(\ref{barIs}) after a somewhat long but
straightforward calculation, where one has to make use of 
the linear relations between the
functions $\phi_{c,\sigma}^{(0)}$ and $\phi_{a,\sigma}^{(0)}$, one
obtains the expression (\ref{corr5}) from the main text with
\bseq
\label{C123}
\begin{align}
\label{C1}
C_1=&(\xi j_\infty^2-2f_\infty)
\frac{\sin\pi\n\,\Gamma(1-\al)\Gamma(1-\al-\n)\Gamma(1-\al+\n)}
{2\sqrt{\pi}\Gamma(3/2-\al)}\;,\\
\label{C2}
C_2=&-\frac{2}{V_{--}^{(0)}}
\int_0^{u_0}\frac{g_-du\,u^{1-d}}{f_0(\lambda_+-\lambda_-)}g
\bigg(\frac{1+\xi j^2}{f}-f\bigg)
\phi_{c,-}^{(0)}\phi_{a,+}^{(0)}\notag\\
&+(\xi j_\infty^2-2f_\infty)\bigg[
\frac{u_0^{2-2\al}}{2\n(1-\al)}
-\frac{V_{+-}^{(0)}}{V_{--}^{(0)}}
\frac{u_0^{2-2\n-2\al}}{2\n(1-\n-\al)}\bigg]\;,\\
\label{C3}
C_3=&2^{-2\n}\frac{\Gamma(-\n)}{\Gamma(\n)}\bigg\{
-\frac{V_{+-}^{(0)}}{V_{--}^{(0)}}C_2
-\frac{\big|V^{(0)}\big|}{\big(V_{--}^{(0)}\big)^2}
\int_0^{u_0}\frac{g_-du\,u^{1-d}}{f_0(\lambda_+-\lambda_-)}
g\bigg(\frac{1+\xi j^2}{f}-f\bigg)
\big(\phi_{a,+}^{(0)}\big)^2\notag\\
&+(\xi
j^2_{\infty}-2f_\infty)\bigg[\frac{u_0^{2+2\n-2\al}}{4\n(1+\n-\al)}
-\frac{V_{+-}^{(0)}}{V_{--}^{(0)}}
\frac{u_0^{2-2\al}}{2\n(1-\al)}
+\bigg(\frac{V_{+-}^{(0)}}{V_{--}^{(0)}}\bigg)^2
\frac{u_0^{2-2\n-2\al}}{4\n(1-\n-\al)}
\bigg]\bigg\}\;,
\end{align}
\eseq
where we have restored the explicit dependence on the functions
defining the domain wall background. 
Note that
these expressions are clearly independent of the arbitrary choice of
the normalization point $u_1$. As already noted before, despite the
fact that Eqs.~(\ref{C2}), (\ref{C3}) apparently contain the other
normalization point $u_0$, they are in fact independent of it
within the accuracy adopted in our calculation
as long as $u_0\gg 1$: the
boundary terms from the integrals are canclelled by the
power-law counterterms. Moreover, the corresponding contributions in the
correlator (\ref{corr5}) are actually important only when they
dominate over the $C_1$-term. For the $C_2$-term this happens if
$\al>1$, while for the $C_3$-term --- if $\al>1+\n$. In these cases
the integrals in (\ref{C2}), (\ref{C3}) become convergent and can be
extended to infinity so that the counterterms disappear.

\section{Energy-momentum of the IR brane}
\label{app:IRbrane}

In Sec.~\ref{sec:states} we consider a brane that cuts off the
large-$u$ portion of the spacetime interpolating between the Lifshitz
region and AdS. Here we show that the energy-momentum of such a brane
can be decomposed into a constant negative tension plus a contribution
satisfying the null energy condition (NEC). We will assume that the
brane is located at an orbifold fixed point under a $Z_2$ reflection
symmetry, similarly to the RS construction
\cite{Randall:1999ee}. Then the energy-momentum of the brane is given
by the junction condition (see e.g. \cite{Berezin:1987bc}),
\[
T_{\m\n}^{brane}=\frac{1}{4\pi G}(K_{\m\n}-\gamma_{\m\n}K_\lambda^\lambda)\;,
\]
where $\gamma_{\m\n}$ is the metric induced on the brane and
$K_{\m\n}$ is the brane's extrinsic curvature. The latter is defined
as
\[
K_{\m\n}=-\nabla_\m n_\n\;,
\]
where $n_\n$ is the inward pointing ($n_u<0$) unit normal to the
brane. A straightforward calculation using the form (\ref{metric}) of
the metric yields
\be
\label{Tbrane}
T_{00}^{brane}=-\frac{d-1}{4\pi G}\cdot\frac{Lf^2}{u^2
  g}\bigg|_{u=\Lambda^{-1}}~,~~~~~~
T_{ij}^{brane}=\frac{\delta_{ij}}{4\pi G}\bigg[
\frac{(d-1)L}{u^2
  g}-\frac{Lf'}{ugf}\bigg]\bigg|_{u=\Lambda^{-1}}\;,
\ee
where both expressions must be evaluated at the position of the brane
$u=\Lambda^{-1}$. We see that the energy density of the brane is
negative. This is the common property of the IR branes in the context
of the RS-type models. It can be accommodated by assuming that the
brane has negative tension. Due to the orbifold symmetry accross the
brane this does not lead to instabilities. In our case one can take
any value of the brane tension satisfying the inequality,
\[
\sigma\leq-\frac{d-1}{4\pi G Lg}\bigg|_{u=\Lambda^{-1}}\;.
\]
Subtracting the corresponding contribution from 
(\ref{Tbrane}) one obtains
the residual energy-momentum tensor ${\bar T}^{brane}_{\m\n}$, such
that for any ($(d+1)$-dimensional) null vector $l^M$
\[
{\bar T}_{\m\n}^{brane}\;l^\m l^\n\geq -\frac{(l^i)^2}{4\pi G}
\cdot\frac{Lf'}{u
  gf}\bigg|_{u=\Lambda^{-1}}\;.
\]
As discussed in Sec.~\ref{sec:geometry}, $f(u)$ is a monotonically
decreasing function on the domain wall solution implying that its
derivative is negative. Thus we conclude that
${\bar T}^{brane}_{\m\n}$ satisfies NEC. 

It follows from the same analysis that if we want to cut off the Lifshitz geometry in the UV by introducing a UV brane, this would need to support a NEC-violating energy-momentum tensor. This agrees with the findings of  \cite{Gubser:2008gr}.

\end{document}